\documentclass[
pre,reprint,longbibliography,
 amsmath,amssymb,
 aps,
]{revtex4-2}

\usepackage{bm}
\usepackage{graphicx}
\usepackage{enumitem}   
\usepackage{appendix}
\usepackage[hidelinks]{hyperref}
\usepackage{cleveref}
\usepackage{float}
\usepackage{color}
\usepackage{ulem}
\usepackage{comment}

\begin{document}

\widetext

\preprint{APS/123-QED}

\title{Fick-Jacobs description and first passage dynamics for diffusion in a channel under stochastic resetting}
\author{Siddharth Jain}
\email{sdj.jain@gmail.com}
\affiliation{Harish-Chandra Research Institute, HBNI, Chhatnag Road, Jhunsi, Allahabad (Prayagraj), UP 211019, India}
\author{Denis Boyer}
 \email{boyer@fisica.unam.mx}
 \affiliation{Instituto de F\'isica, Universidad Nacional Aut\'onoma de M\'exico, Ciudad de M\'exico C.P. 04510 M\'exico.}
 \author{Arnab Pal}
 \thanks{Corresponding author}
 \email{arnabpal@imsc.res.in}
\affiliation{The Institute of Mathematical Sciences, CIT Campus, Taramani, Chennai 600113, India \& 
Homi Bhabha National Institute, Training School Complex, Anushakti Nagar, Mumbai 400094,
India}
\author{Leonardo Dagdug}
 \email{dll@xanum.uam.mx}
\affiliation{Physics Department, Universidad Aut\'onoma Metropolitana-Iztapalapa, San Rafael Atlixco 186, Ciudad de M\'exico, 09340, M\'exico.}

\date{\today}

\begin{abstract}
   Transport of particles through channels is of paramount importance in physics, chemistry and surface science due to its broad real world applications.  
   Much insights can be gained by observing the transition paths of a particle through a channel and collecting statistics on the lifetimes in the channel or the escape probabilities from the channel. In this paper, we consider the diffusive transport through a narrow conical channel of a Brownian particle subject to intermittent dynamics, namely, stochastic resetting. As such, resetting brings the particle back to a desired location from where it resumes its diffusive phase. To this end, we extend the Fick-Jacobs theory of channel-facilitated diffusive transport to resetting-induced transport. Exact expressions for the conditional mean first passage times, escape probabilities and the total average lifetime in the channel are obtained, and their behaviour as a function of the resetting rate are highlighted. It is shown that resetting can expedite the transport through the channel -- rigorous constraints for such conditions are then illustrated. Furthermore, we observe that a carefully chosen resetting rate can render the average lifetime of the particle inside the channel minimal. Interestingly, the optimal rate undergoes continuous and discontinuous transitions as some relevant system parameters are varied. The validity of our one-dimensional analysis and the corresponding theoretical predictions are supported by three-dimensional Brownian dynamics simulations. We thus believe that resetting can be useful to facilitate particle transport across biological membranes -- a phenomena that can spearhead further theoretical and experimental studies.  
\end{abstract}

\maketitle

\section{Introduction}

The problem of particle transport through confined geometries containing narrow openings and bottlenecks has recently led to many theoretical and experimental research fronts. In his original work dating back to 1855, Adolf Fick was the first scientist to treat diffusion into a channel of varying cross-sections as a one-dimensional system \cite{fick}. Later, the seminal works by Jacobs and Zwanzig triggered renewed research on this topic. The so-called Fick-Jacobs approach consists of eliminating transverse stochastic degrees of freedom by assuming fast equilibration in such directions \cite{Jacobs,Zw,Szabo}. As Zwanzig pointed out, the key quantity that characterizes the unbiased diffusive transport of point-like Brownian particles in quasi-one-dimensional systems with periodically varying geometric constraints is the effective diffusivity \cite{Zw}. As a result, enormous theoretical efforts have been made to obtain consistent position-dependent diffusion coefficients that rely upon the exact nature of confinement \cite{Zw,Jacobs,Szabo,BDB2015,RR,KPre,MSP,CCD,SD,BDB2017,Burada,KP,DP,Pompa}.  

Diffusion of particles, molecules, or even living microorganisms in confined geometries such as tubes and channels plays a key role across various scales in natural and technological processes \cite{JMR,Haul,Han,Gershowand,Hille,DD,Keyser,ref-11,ref-12,ref-13}. It is now well understood that membrane transport of metabolites and other solutes is facilitated by membrane proteins that form water filled channels \cite{Rostovtseva,Hoogerheide}. The basic mechanism of channel-regulated transport has thus been a focal point of both theoretical and practical research. However, despite many years of studies, new phenomena related to channel-facilitated metabolite transport continue to be found and a comprehensive understanding is still to be developed.

A diffusing particle inside a channel can either escape from both sides of the membrane, or it may be allowed to escape through one side while the other side simply reflects it back. Key quantities in these processes are the first passage probabilities and average escape times from
the channel (or lifetimes) conditioned to the exit side of the membrane, as well
as the overall average lifetime of the particle in the channel. These quantities are ubiquitous in theoretical studies on channel-facilitated transport \cite{Berezh06,Dagdug-03}. Remarkably, these observables have also taken center stage since recent advances in single-particle experiments enable direct observation of transition or first passage paths, marking the events when a molecule diffuses across an activation barrier or reacts to form products \cite{BCMSV}. Fluorescence and force spectroscopy methods render observations which hold treasure trove of information on mechanisms such as molecular dynamics, structures of chemical networks or biological membrane channels \cite{Alex20,Chung12}. Moreover, single molecule experiments have unraveled the salient roles played by the channel shape, the dynamics of biomolecular folding, binding or catalyzing, as well as by the interactions between the channel and the molecule -- all of which pave the way to our understanding of membrane transport of metabolites or ions. Naturally, the study of first passage distributions is crucial to understand how the biomolecules fulfil their function, e.g., fold, bind, or react inside a membrane channel.

Indeed, first-passage time processes have overarching applications in physics, chemistry, biology, computer science, ecology and
other cross-disciplinary fields \cite{Redner}. From macroscopic search processes where a group of searchers or agents look for
some resources or animals forage to find prey or wild resources \cite{BCMSV} to microscopic search where a regulatory
protein has to find a target sequence on the DNA, or chemical and biochemical reactions \cite{HTB,CBVM}, trafficking receptors on biological membranes \cite{HolSchu} -- these are all canonical examples of stochastic first passage processes. Processes such as the spreading of sexually transmitted diseases in a human social network or of viruses across  the world wide web \cite{GSHM} are also controlled by first encounter events \cite{BV}. Recently, first passage processes which are further subject to resetting have gained immense interest due to their myriad of applications in statistical physics \cite{review,Restart1,Restart2,Pal-potential,transport1,transport2,Pal-time-dep,sth-1,eco-2,ctrw,kpz,TS,review-2}, chemical process \cite{Reuveni-14}, biology \cite{Restart-Bio-0,Restart-Bio-1,Restart-Bio-2,Restart-Bio-3}, queuing theory \cite{q5}, and computer science \cite{Luby,algorithm,algorithm-2,Montanari}. In these problems, a stochastic variable of interest ({\it e.g}, the position of a diffusing particle) is brought back from time to time to a specific position, from which it restarts anew. In Michaelis-Menten enzyme-catalyzed reactions, unbinding can be viewed as a resetting event \cite{Reuveni-14}. Furthermore, the sudden falls
in stock market price, the sharp layoff of individual jobs due to post-pandemic recession, or the massive extinction of population due to catastrophes are also canonical signatures of resetting events \cite{jobs-r,population-r}. A hallmark property of resetting is its ability to speed-up complex search processes, motivating a flurry of theoretical \cite{review,Restart1,Restart2,HRS,CV-c9,CV-c11,kusmierzprl,ReuveniPRL,PalReuveniPRL,palprasad,landaupal,partial-bc,SR-1,SR-2,ex-1,reset-networks} and experimental works \cite{expt-1,expt-2}. 

In this work, we attempt to study channel-facilitated first passage transport in the presence of resetting. First passage properties under resetting have been studied extensively for diffusive and active systems in simple one-dimensional intervals with two absorbing/reactive boundaries \cite{palprasad,landaupal,partial-bc,ahmad}, or one reflective-one absorbing boundary \cite{Restart-Bio-2,ahmad,freezing}. However, to the best of our knowledge, diffusion in  $2D$ or $3D$ channels of varying width in the presence of resetting has not been explored so far. While the lifetime statistics of reset-diffusing particles inside the channel is of interest from the perspective of transport, another interesting question that naturally emerges is whether such transport can be made optimal through resetting. Furthermore, while it has been shown that diffusion-controlled reactions of solutes with a reaction site hidden within a protein membrane channel or a protein cavity can be slowed down by many orders of magnitude \cite{Dagdug-03}, resetting or unbinding events can expedite on the contrary the turnover rate of enzymatic catalysis process \cite{Reuveni-14}. Thus, the combination of these two phenomena -- resetting and geometric confinement -- can be a reasonable attempt to delve deeper into the understanding of Michaelis-Menten when the reaction site is hidden in a protein membrane channel or protein cavity.

\begin{figure}[!ht]
\centering
\includegraphics[width=0.4\textheight]{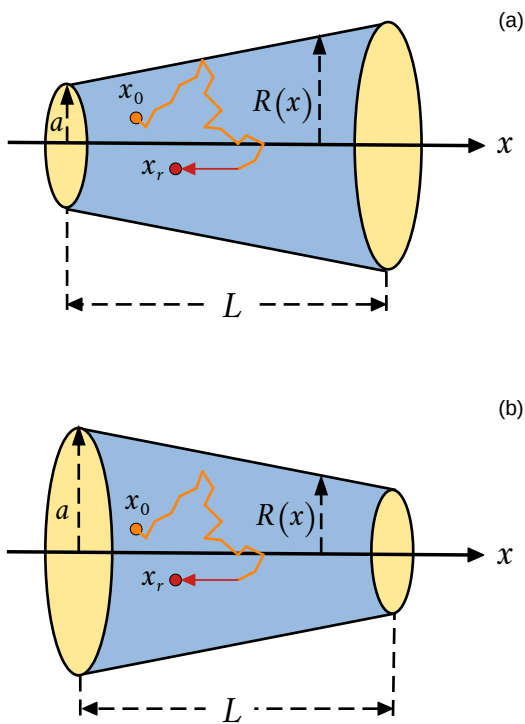}
\caption{Schematic representation of a three-dimensional conical tube of variable radius $R(x)$ and length $L$ with a constant radius variation rate, $|dR(x)/dx | = const = \lambda$. In this case, the effective diffusivity is position-independent and a function of $\lambda$ through the relation $D_\lambda = D_0 / \sqrt{1+\lambda^2}$. Panels (a) and (b) show an expanding and narrowing tube, respectively. The particle starts from the initial position at $x_0$ and is reset to a fixed position $x_r$ at a rate $r$, from where it resumes diffusion until the next reset. The channel ends at $x=0$ and $x=L$ (shown in yellow) could be either reflecting or absorbing.}
\label{fig:cone}
\end{figure}

As mentioned before, particle transport inside a channel can well be captured by the Fick-Jacobs (FJ) formalism. In here, we extend this formalism for diffusing particles under resetting. General expressions for the escape probabilities and the
average life times for the particles escaping through each boundary are computed. These lifetimes are conditional because they are calculated for the two subset of all possible realizations of the escape process. The total average lifetime of the particle in the channel is then obtained as a weighted sum of these conditional lifetimes where the escape probabilities are used. To illustrate some qualitative features of the general theory, we study a special case in which the particle diffuses through a three dimensional conical channel (see Fig. \ref{fig:cone}). Moreover, we consider that the particle is stochastically reset back to a pre-defined location at a given rate. We find that resetting can expedite or slow down the escape of the particle from the channel. Utilizing the resetting criterion when the resetting and starting positions coincide, we find the exact conditions conferred by the system parameters such as the initial position, channel length and channel slope that determine when resetting is going to be beneficial. In this case, we obtain the optimal condition when the escape time can be minimized. Interestingly, we find that the optimal resetting rate can undergo continuous or discontinuous transitions with varying the initial position as the control parameter. Furthermore, an intriguing optimization is also observed for the conditional times. Our theoretical results are corroborated with Brownian dynamics simulations in 3D channels.

The structure of the paper is as follows. We describe the model setup and review the Fick-Jacobs formalism in Section (\ref{model}). We then introduce the resetting dynamics and sketch out the steps leading to the modified Fick-Jacobs formalism in Section (\ref{FJ-reset-renewal}). Therein, we provide general expressions and solutions for the observables of interest. In Section (\ref{cone-two-abs}), we discuss the first passage statistics of the particle in a three dimensional cone with two absorbing boundaries at the end. We discuss the effects of resetting, including emergence of an optimal resetting rate.  The following section (\ref{sec:reflabs}) is dedicated to the study of first passage statistics in a three dimensional cone with one absorbing boundary and one reflecting boundary at the other end. Some of the results have been moved to the Appendix for brevity. We summarize our paper and put concluding remarks in Section (\ref{conclusions}).

\section{Model details and formalism}

\subsection{Setup and Fick-Jacobs equation}
\label{model}
We consider diffusion of a particle inside a three dimensional conical tube of variable radius $R(x)$ and length $L$. The radius of the tube linearly grows in space (see Fig. \ref{fig:cone}). When diffusion occurs in quasi-one-dimensional structures, one can map particle motion onto an effective one-dimensional (1D) description in terms of diffusion along the channel axis. The key point of the derivation is the assumption of equilibration in the transverse direction. When the system geometry is smoothly varying along the $x$-coordinate measured along the tube/channel axis, the simplest one-dimensional description is given by the Fick-Jacobs equation \cite{Jacobs}. In terms of the probability distribution $c(x,t)$,
the FJ equation reads
\begin{equation}\label{eq:FJ}
\frac{\partial {c}(x,t)}{\partial t}=D_0 \frac{\partial}{\partial x}\left\{{A}(x)\frac{\partial}{\partial x}\left[\frac{{c}(x,t)}{{A}(x)} \right]\right\},
\end{equation}
where $D_0$ is the bare diffusion constant, $A(x)$ is the cross-section area $\pi R^2(x)$ of a tube of radius $R(x)$, or the width $w(x)$ of a two-dimensional channel.

It has been shown that a more accurate reduction to the 1D description results in a position-dependent diffusivity $D(x)$. Then the probability density in the channel satisfies a modified Fick-Jacobs equation which was first derived by Zwanzig \cite{Zw}
\begin{equation}
\frac{\partial {c}(x,t)}{\partial t}=\frac{\partial}{\partial x}\left\{D(x){A}(x)\frac{\partial}{\partial x}\left[\frac{{c}(x,t)}{{A}(x)} \right]\right\}.
\label{FJZw}
\end{equation}
Confinement in higher dimensions gives rise to an effective entropic potential in the reduced dimension. In fact, Eq. \ref{FJZw} is formally equivalent to the Smoluchowski equation
\begin{equation}
\frac{\partial {c}(x,t)}{\partial t}=\frac{\partial}{\partial x}\left\{D(x)e^{-\beta U(x)}\frac{\partial}{\partial x}\left[e^{\beta U(x)} c(x,t) \right] \right\},
\label{Sm}
\end{equation}
where the entropy potential is given by -$\beta U(x)=\ln(w(x)/w(x_0))$ for a 2D channel and -$\beta U(x)=\ln(R^2(x)/R^2(x_0))$ for a 3D tube. Here, $\beta = 1/(k_B T)$ is the inverse temperature with $k_B$ the Boltzmann constant.

There are several interpretations for the position dependent effective diffusion coefficient. For example, in the case of a narrow 2D channel of varying width, it was shown by Reguera and Rubi \cite{RR} that the following relation holds 
\begin{equation}
D(x)\approx D_{RR}^{2D}(x) = \frac{D_0}{\left[1 +  \frac{1}{4} w'^2(x)\right]^{\eta}},
\label{eq:RR2D}
\end{equation}
where $w'(x)=dw(x)/dx$ and $D_0$ the bare diffusion coefficient. The index $\eta$ takes the value $1/3$ for the 2D set-up. Parallel derivations were given by Kalinay and Percus \cite{KP}, Martens \textit{et al}. \cite{MSP} and Garc\'ia-Chung  and co-workers \cite{CCD}. 
In a similar manner, the following spatial dependencies were obtained for the 3D tube geometry by Reguera and Rubi \cite{RR} and later by Kalinay and Percus \cite{KP}
\begin{equation}
D(x)\approx D_{RR}^{3D}(x) = D_{KP}^{3D}(x) = \frac{1}{\sqrt{1+ R'(x)^2}}D_0,
\label{eq:RR3D}
\end{equation}
where $\eta=1/2$ was assumed for 3D set-up. Here we consider 3D conical tube so that the entropic potential satisfies $e^{-\beta U(x)}=R^{2}(x)/R^{2}(x_0)$, as mentioned before. Furthermore, we will consider a linear variation of the tube radius. In particular, the tube radius $R(x)$ increases in the axial direction with a constant rate $\lambda$ so that $R(x)=a+\lambda x$, where the $x$-coordinate is measured along the tube axis, and $a$ is the tube radius at $x=0$.  It then follows from Eq. \eqref{eq:RR3D} that the effective diffusivity is position-independent, hence the Reguera-Rubi
formula reduces to 
\begin{align}
D_\lambda=\frac{D_0}{\sqrt{1+\lambda^2}},    
\label{diffusion-constant}
\end{align}
which we will use in this study unless otherwise stated.

\subsection{Fick-Jacobs equation with resetting and escape statistics}
\label{FJ-reset-renewal}
We now consider the same point-like Brownian particle diffusing in a spatially constrained asymmetric 3D tube, but it is now intermittently reset to the position $x_r$ at a constant rate $r$. Let us denote the position distribution for this particle as $c_r(x,t|x_0,0)$, where suffix $r$ indicates resetting. This position density is not normalized to unity but to the survival probability $Q_r(x_0,x_r,t)$ since the particles will eventually escape through the boundaries. $Q_r(x_0,x_r,t)$ is defined as the probability that the particle has stayed in-between the boundary coordinates up to time $t$, having started from $x_0$ and with resetting at $x_r$,
\begin{align}
   Q_r(x_0,x_r,t)=\int_0^L dx ~c_r(x,t|x_0,0).
\end{align}
A similar relation holds also for the resetting-free process, {\it i.e.}, $Q(x_0,t)=\int_0^L~dx~c(x,t|x_0,0)$.

After each resetting event, the particle returns to the same location $x_r$ and restarts its motion, keeping no memory of previous resetting events. Thus the process is renewed after each resetting and we can write a time dependent equation for the position density $c_r(x,t|x_0,0)$ using a renewal formalism \cite{palprasad}
\begin{align}
c_r(x,t|x_0,0) &= e^{-rt}c(x,t|x_0,0) \nonumber \\
&+ r\int_0^t~d\tau~e^{-r (t-\tau)}~c(x,t|x_r,\tau) Q_r(x_0,x_r,\tau)~,
\label{propagator-renewal}
\end{align}
where the first term indicates that there was no resetting event up to a time $t$. On the other hand, the second term captures a physical scenario when the particle undergoes multiple resetting to $x_r$ not being absorbed by the boundaries up to time $\tau$, followed by a resetting at time $\tau$, and a last excursion without resetting (starting from $x_r$) of duration $t-\tau$.  This equation clearly states that the concentration in the presence of resetting depends on the concentration of the underlying resetting-free process and on the survival probability of the reset process. We can then write an effective Fick-Jacobs equation in the presence of resetting, which reads
\begin{align}
    \frac{\partial {c_{r}}(x,t)}{\partial t}= & \frac{\partial}{\partial x}\left\{D(x)e^{-\beta U(x)}\frac{\partial}{\partial x}\left[e^{\beta U(x)} c_{r}(x,t) \right] \right\} \nonumber\\
    & -r c_{r}(x,t|x_0) + r \delta(x-x_r)Q_r(x_0,x_r,t).
\label{Smr}
\end{align}
The effective FJ equation has the usual diffusive and drift terms in addition to new two terms. While the first one represents the removal/outflux of the particle from its current position $x$ at rate $r$, the second one indicates a source term only at the resetting position where the probability is being accumulated. This is however conditioned on the particle's survival -- hence the appearance of the survival probability $Q_r$ in the last term.

%

We need not solve these resetting equations explicitly. To obtain the survival probability in the presence of resetting, we notice that
one can similarly write a renewal equation for this quantity \cite{review,Pal-time-dep},
\begin{align}
    Q_r(x_0,x_r,t)&=e^{-rt} Q(x_0,t) \nonumber \\
  &+\int_0^t~d\tau~re^{-r\tau}~Q(x_0,\tau)~ Q_r(x_r,x_r,t-\tau),
    \label{ND-1}
\end{align}
where $Q(x_0,t)$ denotes the survival probability of the resetting-free process.
Taking Laplace transform on both sides of Eq. (\ref{ND-1}) we arrive at the following relation
\begin{align}
    q_r(x_0,x_r,s)&=\frac{q(x_0,s+r)}{1-rq(x_r,s+r)} ,
    \label{ND-LT}
\end{align}
where $q_r(x_0,x_r,s)=\int_0^{\infty}dt\ e^{-st}Q_r(x_0,x_r,t)$ and $q(x_0,s)=\int_0^{\infty}dt\ e^{-st}Q(x_0,t)$. Similarly,
taking the Laplace transform on both sides of Eq. (\ref{propagator-renewal}) and using Eq. (\ref{ND-LT}), we obtain
\begin{align}
    \tilde{c_r}(x,s|x_0,0)&=\tilde{c}(x,s+r|x_0,0)\nonumber \\
    &+\frac{rq(x_0,s+r)\tilde{c}(x,s+r|x_r,0)}{1-r q(x_r,s+r)}~,
\label{propagator-formula}
\end{align}
which is a crucial relation relating the concentration in the presence of resetting only in terms of the observables of the underlying process. 

Since we are interested in the escape properties of the particle from the channel, we need to examine the first passage time statistics. Generically, this encodes the information for the lifetime of the particle in confinement, and the first escape is marked with suitable boundary conditions. The first passage time density of any process is related to the survival probability through the general relation \cite{Redner}
\begin{align}
    f_r(t|x_0,x_r)=-\frac{\partial Q_r(x_0,x_r,t)}{\partial t},
\end{align}
which in the Laplace domain translates to $\tilde{f_r}(s|x_0,x_r)=1-sq_r(x_0,x_r,s)$. In the following we will consider the case where the resetting position is the starting position, or $x_r=x_0$, unless otherwise stated. Similar to the survival probability, the first passage time densities with and without resetting are also related in the Laplace domain. From Eq. (\ref{ND-LT}),
\begin{align}
    \tilde{f}_{r}(s|x_0)=\frac{(s+r)\tilde{f}(s+r|x_0)}{s+r\tilde{f}(s+r|x_0)}.
    \label{firstpassage-renewal}
\end{align}
The mean first passage time (MFPT) is simply given by
\begin{align}\label{MFPTGen}
    \langle T_r(x_0) \rangle=\int_0^\infty dt ~t f_r(t|x_0,x_r)= q_r(x_0,x_0,s \to 0).
\end{align}
Recall that within the current setup, the particle starting from $x_0$, can escape through any of the boundaries. This is captured by the unconditional MFPT above. In other words, it measures the average lifetime of the particle inside the channel.
However, one can also look for the transit time i.e., the first passage times that estimate the time when the particle escapes through one designated boundary. These are conditional times and have markedly different statistics than the MFPT.

The conditional mean transit times (in the following, `+' implies observables related to the right boundary at $L$ and `-' implies the same for the left boundary at $0$) can be computed from the current/flux conditioned on respective boundaries, {\it i.e.}, 
\begin{align}
\langle \tau_r(x_0) \rangle^{\pm}=\frac{\int_0^\infty dt~t~J_r^{\pm}(x_0,t)}{\int_0^\infty dt~J_r^{\pm}(x_0,t)},
\label{mean-exit-times-definition}
\end{align}
where $J_r^{+}(x_0,t)$ $[J_r^{-}(x_0,t)]$ is the probability flux to the right(L) [left(0)] boundary respectively. The terms in the denominator 
correspond to the exit/splitting probabilities through the boundary $L$ and $0$ respectively
\begin{align}\label{eq:epdplus}
    \epsilon_r^\pm(x_0)=\int_0^\infty~dt~J_r^\pm(x_0,t)= j_r^{\pm}(x_0,s=0),
\end{align}
where $j_r^{\pm}(x_0,s)=\int_0^{\infty}dt~e^{-st}~J_r^{\pm}(x_0,t)$ are the Laplace transform of currents.
Note that $\epsilon^+ + \epsilon^-=1$, as the particle eventually escapes the channel.
Eqs. (\ref{mean-exit-times-definition}) can be rewritten as
\begin{align}
\langle \tau_r(x_0) \rangle^{\pm}=\frac{-\frac{\partial j_r^{\pm}(x_0,s)}{\partial s}|_{s \to 0}}{\epsilon_r^{\pm}(x_0)}~.
\label{mean-exit-times-definition-LT}
\end{align}
Note that one should always have $    \langle T_r(x_0) \rangle = \epsilon_r^+ \langle \tau_r \rangle^+ +\epsilon_r^- \langle \tau_r \rangle^- $ for both underlying and reset process. The conditional first passage densities for the corresponding transit times are also related to the currents with proper normalisation \cite{palprasad}
\begin{align}
    f_r^\pm(t|x_0)=\frac{J_r^\pm(x_0,t)}{\epsilon_r^\pm(x_0)}.
    \label{transit-PDF}
\end{align}
While the currents $j_r^{\pm}(x_0,s)$ can be computed from the knowledge of the conditional propagator $\tilde{c_r}(x,s|x_0,0)$, we take a different route 
based on the renewal approach suitable for resetting processes. Technically, we use the renewal relations for the currents and hence the conditional first passage densities for the resetting process. The usefulness of these relations should be appreciated since they connect observables of the process to those of the underlying process. Thus, the task significantly reduces to the computation of the currents and (un)conditional densities for the underlying process. Then one can use the renewal relation to obtain the desired results. A renewal relation for the current can be directly obtained from Eq. (\ref{propagator-formula}) (also see \cite{palprasad} for detailed derivation)
\begin{align}
    j_r^{\pm}(x_0,s)=\frac{j^{\pm}(x_0,s)}{1-rq(x_0,s+r)},
    \label{currents-renewal}
\end{align}
where $j^{\pm}(x_0,s)$ is the current for the underlying process in Laplace space. Using the above and Eq. (\ref{transit-PDF}), we find a similar renewal relation that connects the conditional first passage time densities for the reset process to the same for the underlying process
\begin{align}
\tilde{f}_r^\pm(s|x_0)=\frac{\tilde{f}^\pm(s+r|x_0)}{\tilde{f}^\pm(r|x_0)} \frac{(s+r)\tilde{f}(r|x_0)}{s+r\tilde{f}(s+r|x_0)},
\label{conditionalFPT-renewal}
\end{align}
which will prove to be a useful relation later in our study to compute the moments $\langle \tau_r(x_0) \rangle^\pm$ for the conditional times. 


\subsection{General solution of FJ equation}
To make use of the renewal relations, one first needs to obtain the complete statistics for the underlying process. In this section, we provide solutions for the propagator which will serve as a backbone for computing currents, survival and first passage time density for the underlying reset free processes.
To this end, let us first recall the Smoluchowski equation from Eq. (\ref{Sm})
\begin{eqnarray}
\frac{\partial {c}(x,t)}{\partial t}= & \frac{\partial}{\partial x}\left\{D(x)e^{-\beta U(x)}\frac{\partial}{\partial x}\left[e^{\beta U(x)} c(x,t) \right] \right\},
\end{eqnarray}
and the modified Fick-Jacobs equation for the 3D cone (see Fig.\ref{fig:cone}) \begin{align}
\frac{\partial {c}(x,t)}{\partial t}=&  D_\lambda\frac{\partial}{\partial x}\left\{ R^2(x)\frac{\partial}{\partial x} \left[\frac{{c}(x,t)}{R^2(x)} \right]\right\}. \label{propagator}
\end{align}
To proceed further, we first scale the probability density by the radius of the channel, namely,
\begin{align}
    c(x,t)=R(x)g(x,t),
    \label{sep-var}
\end{align}
and define the following Laplace transforms 
\begin{align}
    \Tilde{c}(x,s|x_{0},0)&=\int_{0}^{\infty} c(x,t|x_{0},0) ~e^{-st}dt, \\
        G(x,s)&=\int_{0}^{\infty} g(x,t) ~e^{-st}dt,
\end{align}
so that Eq. (\ref{FJZw}) is Laplace transformed as
\begin{align}
    sR(x)G(x,s)-\delta (x-x_{0})=\frac{\partial}{\partial x}\left\{D_\lambda R^{2}(x)\frac{\partial}{\partial x}\left[\frac{G(x,s)}{R(x)} \right]\right\}.
\label{FJZw-2}
\end{align}
We consider the conical case $R(x)=\lambda x+a$, where $\lambda\ge0$ for an expanding channel and $\lambda\le0$ for a shrinking one. Eq. \eqref{FJZw-2} becomes
\begin{align}\label{condderx0}
\frac{-\delta (x-x_{0})}{\lambda x_{0}+a}=D_\lambda\frac{\partial ^{2} G}{\partial x^{2}}-sG.
\end{align}
The general solution to the above equation on each side of $x_0$ is given by
\begin{align}
    G(x,s)=c_{1}e^{-\sqrt{s/D_\lambda}x}+c_{2}e^{\sqrt{s/D_\lambda}x} , \quad \quad x \in  [0,x_{0}) \\
    G(x,s)=c_{3}e^{-\sqrt{s/D_\lambda}x}+c_{4}e^{\sqrt{s/D_\lambda}x} ,\quad \quad x \in  (x_{0},L] 
\end{align}
resulting in
\begin{align}\label{eq:sect}
    \tilde{c}(x,s)=(\lambda x+a)[c_{1}e^{-\sqrt{s/D_\lambda}x}+c_{2}e^{\sqrt{s/D_\lambda}x}] , \quad x \in  [0,x_{0})
\end{align}
\begin{align}\label{eq:sect1}
    \tilde{c}(x,s)=(\lambda x+a)[c_{3}e^{-\sqrt{s/D_{\lambda}}x}+c_{4}e^{\sqrt{s/D_\lambda}x}] , \quad x \in  (x_{0},L] .
\end{align}
To find the four constants we need to impose one condition at each boundary which we will discuss in the following. The additional two conditions are obtained from matching the propagator, namely, 
$(i)$ $ \tilde{c}(x,s)$ is continuous at $x=x_0$, and $(ii)$ the derivative  $\partial_x\tilde{c}(x,s)$ is discontinuous at $x=x_{0}$, as seen from integrating 
Eq. (\ref{condderx0}) over a narrow interval around $x_0$,
\begin{equation}
    \left.\frac{\partial G}{\partial x}\right|_{x_0^+} - \left.\frac{\partial G}{\partial x}\right|_{x_0^-}=-\frac{1}{D_{\lambda}(\lambda x_0+a)}.
\end{equation}

\section{First passage observables in a three dimensional cone with two absorbing boundaries}
\label{cone-two-abs}
We first consider the conical tube
with two absorbing points at $x=0,L$. As stated before, it has a radius which varies spatially with a rate $\lambda$ and the effective diffusive constant is $D_\lambda$ given by Eq. (\ref{diffusion-constant}). By symmetry, it is sufficient to consider $\lambda\ge0$.
The boundary conditions read
\begin{align}
    c(0,t)=0~,~c(L,t)=0,
\end{align}
since a particle will be immediately absorbed if it reaches one of the boundaries. The particle starts from an initial position $x_0$ within the cone and it is reset to the same point at a rate $r$.
Using the matching and boundary conditions, we solve Eqs. (\ref{eq:sect}) and (\ref{eq:sect1}) to obtain the exact expressions for the constants \begin{gather}
    c_{1}=\frac{-\sinh(u(L-x_{0}))}{2uD_\lambda(\lambda x_{0}+a)\sinh(uL)}, \\
    c_{2}=\frac{\sinh(u(L-x_{0}))}{2uD_\lambda(\lambda x_{0}+a)\sinh(uL)}, \\
    c_{3}=\frac{e^{uL}\sinh(ux_{0})}{2uD_\lambda(\lambda x_{0}+a)\sinh(uL)}, \\
     c_{4}=\frac{-e^{-uL}\sinh(ux_{0})}{2uD_\lambda(\lambda x_{0}+a)\sinh(uL)},
\end{gather}
where $u=\sqrt{s/D_\lambda}$.

\subsection{Currents through the boundaries}
The probability flux/current entering the left (right) absorbing boundary at $x=0$ ($L$, resp.) for the resetting free process can be found by observing at the modified FJ equation in Eq. (\ref{propagator}) which can be written in the form of a continuity equation $\frac{\partial c}{\partial t}=-\frac{\partial J}{\partial x}$, where one identifies $J$ to be the probability current. Then, the probability flux through the right boundary can be written as
\begin{align}
    J^+(x_0,t)=
    -D_\lambda R^{2}(x)\frac{\partial}{\partial x}\left[\frac{c(x,t)}{R^{2}(x)} \right]{\bigg|_{x=L}} 
   &=
-D_\lambda\frac{\partial c}{\partial x}\bigg|_{x=L},
\label{right-J-r=0}
\end{align}
where we have used $c(x,t)=0$ at the boundary $L$.
A similar analysis holds for the left boundary, where the probability flux/current is given by
\begin{align}
    J^-(x_0,t)=D_\lambda R^{2}(x)\frac{\partial}{\partial x}\left[\frac{{c}(x,t)}{R^{2}(x)} \right]\bigg|_{x=0}= 
D_\lambda\frac{\partial c}{\partial x}\bigg|_{x=0},
\label{left-J-r=0}
\end{align}
where again we have used the boundary condition at $x=0$. The integral of the fluxes in time from zero to infinity gives the splitting probability $    \epsilon^\pm(x_0)=\int_0^\infty~dt~J^\pm(x_0,t)$. Similarly, the conditional transit time densities read 
$    f^\pm(t|x_0)=J^\pm(x_0,t)/\epsilon^\pm(x_0)$ [see Eq. (\ref{transit-PDF})]. These expressions will be used in the next section.

\begin{figure*}[ht!]
		\includegraphics[width=7.5cm]{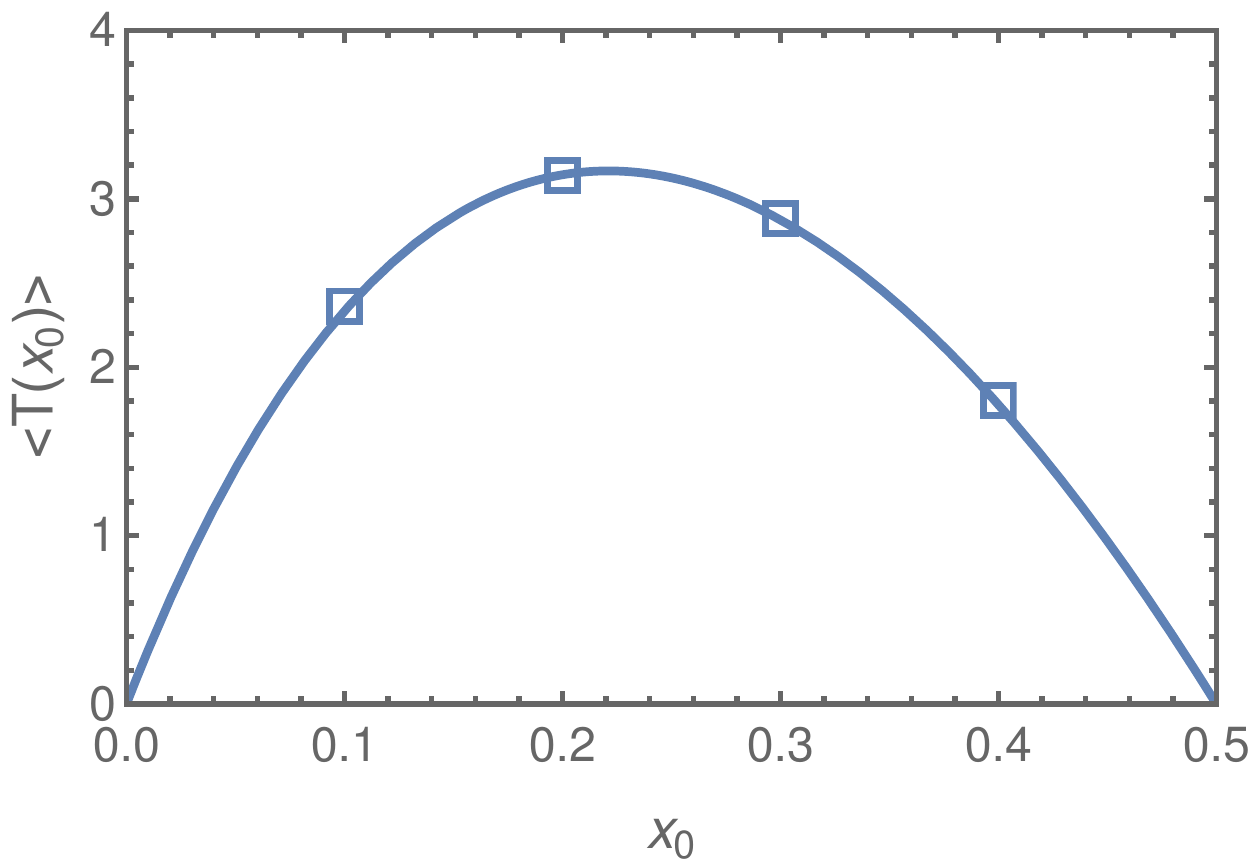}
		\includegraphics[width=9cm]{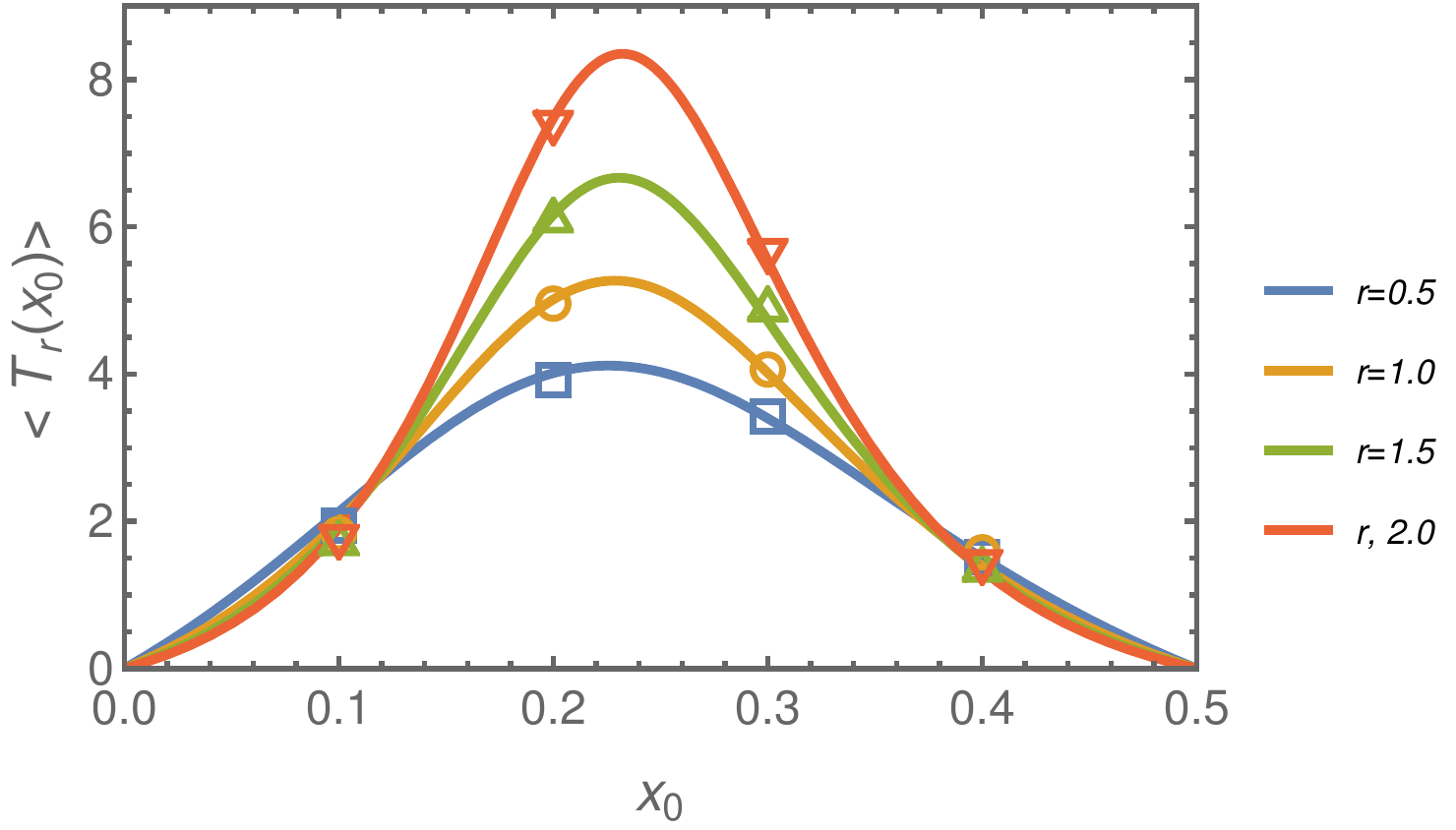}
	\caption{Variation of MFPTs as a function of $x_0$ without resetting (left) and with resetting (right) in the absorbing-absorbing case predicted by Eq. \eqref{T2abs} (solid lines). The escape times obtained from Brownian dynamics simulations are shown with open symbols ($a=0.1$, $D_\lambda=0.01$, $L=0.5$, $\lambda$=0.2).}
	\label{fig:pos_abs}
\end{figure*}


\subsection{Unconditional and conditional first passage time for the underlying process}
Since a particle can get absorbed at either of the two boundaries, the total flux will be the sum of these fluxes given by Eqs.  (\ref{right-J-r=0}) and (\ref{left-J-r=0}). Thus, the unconditional first passage density from either boundary is given by
\begin{align}
    f(t|x_{0}) = J_+(x_0,t)+J_-(x_0,t).
\end{align}
Laplace transforming the above leads to
\begin{align}\label{eq:fptd}
    \tilde{f}(s|x_{0})=j^{+}(x_{0},s)+j^{-}(x_{0},s),
\end{align}
where each term can be computed easily from the propagator $\tilde{c}(x,s)$ and using Eqs. (\ref{right-J-r=0}) and (\ref{left-J-r=0}). One gets
\begin{align}\label{eq:jp}
   j^{+}(x_0,s)&=-D_\lambda \frac{\partial \tilde{c}}{\partial x}\bigg|_{x=L}
  =-D_\lambda[R(x)\frac{\partial G}{\partial x}+ G \lambda]\bigg|_{x=L} \nonumber\\
&=(\lambda L+a)\frac{\sinh(ux_{0})}{(\lambda x_{0}+a)\sinh(uL)},
\end{align}
and
\begin{align}\label{eq:jm}
     j^{-}(x_0,s)&=D_\lambda \frac{\partial \tilde{c}}{\partial x}\bigg|_{x=0}  =D_\lambda[R(x)\frac{\partial G}{\partial x}+ G \lambda]\bigg|_{x=0} \nonumber\\
&=a\frac{\sinh(u(L-x_{0})))}{(\lambda x_{0}+a)\sinh(uL)},
\end{align}
while going from the first to the second line we have used $\tilde{c}(x,s)=R(x)G(x,s)=(a+\lambda x)G(x,s)$. The unconditional first passage density is then given in Laplace space by
\begin{align}\label{eq:fsaa}
   \tilde{f}(s|x_0)=\frac{(\lambda L+a)\sinh(ux_{0})+a \sinh(u(L-x_{0}))}{(\lambda x_{0}+a)\sinh(uL)}.
\end{align}
 The moments of the first passage time distribution can be obtained from the small $s$ expansion of the quantity $\tilde{f}(s|x_0)\simeq \int_0^{\infty}dt f(t|x_0) (1-st+\frac{s^2}{2}t^2+...)$. In terms of the variable $u=\sqrt{s/D_\lambda}$, one has:
\begin{equation}
    \tilde{f}(s|x_0)=1-D_\lambda \langle T(x_0)\rangle u^2 +\frac{D_\lambda^2\langle T^2(x_0)\rangle}{2} u^4+...
\end{equation}
By expanding Eq. (\ref{eq:fsaa}) up to 4th order in $u$, one can obtain the mean first passage time and the second moment for the underlying process, respectively, as
\begin{align}\label{eq:Taa}
    \langle T(x_{0})\rangle=\frac{x_{0}(L-x_{0})(3a+\lambda (L+x_{0}))}{6D_\lambda(\lambda x_{0}+a)},
\end{align}
and
\begin{widetext}
\begin{align}\label{eq:T2aa}
    \langle T^{2}(x_{0})\rangle=\frac{x_{0}(L-x_0)\left[15a(L^{2}+Lx_{0}-x^{2}_{0})+\lambda(x_0+L)(7L^2-3x_0^2) \right]}{180D_\lambda^{2}(\lambda x_{0}+a)}.
\end{align}
\end{widetext}
Note that the moments can also be computed directly from Eq. (\ref{eq:fsaa}) by using the formula $\langle T^n \rangle=(-1)^n \frac{d^n}{ds^n}\tilde{f}(s|x_0)\bigg|_{s \to 0}$. The conditional first passage time or exit time densities can be computed from the conditional threshold currents (see Eq. (\ref{transit-PDF}))
\begin{align}
    \tilde{f}^{+}(s|x_0)&=\frac{L\sinh(ux_0)}{x_0\sinh(uL)}, \nonumber\\
     \tilde{f}^{-}(s|x_0)&=\frac{L\sinh(u(L-x_0))}{(L-x_0)\sinh(uL)}.
     \label{conditionalFPT-LT-underlying}
\end{align}
From these densities and using Eq. (\ref{mean-exit-times-definition-LT}), we obtain the mean conditional exit times
\begin{align}
    \langle \tau(x_0) \rangle^{+} &= \frac{L^2-x_{0}^{2}}{6D_{\lambda}}, \nonumber \\
    \langle \tau(x_0) \rangle^{-} &= \frac{2Lx_0-x_{0}^{2}}{6D_{\lambda}}.
\label{ctm}    
\end{align}
The corresponding splitting probabilities are given by
\begin{align}
    \epsilon^{+}(x_0)&=j^{+}(x_0,s=0)=\frac{x_0(\lambda L+a)}{L(\lambda x_0+a)}, \nonumber \\
    \epsilon^{-}(x_0)&=j^{-}(x_0,s=0)=\frac{a(L-x_0)}{L(\lambda x_0+a)},
\end{align}
 which were also known for $x_0=0$ \cite{BDB2017}.

\subsection{Unconditional first passage time for reset process}
\begin{figure*}[ht!]
		\includegraphics[width=0.45\textwidth]{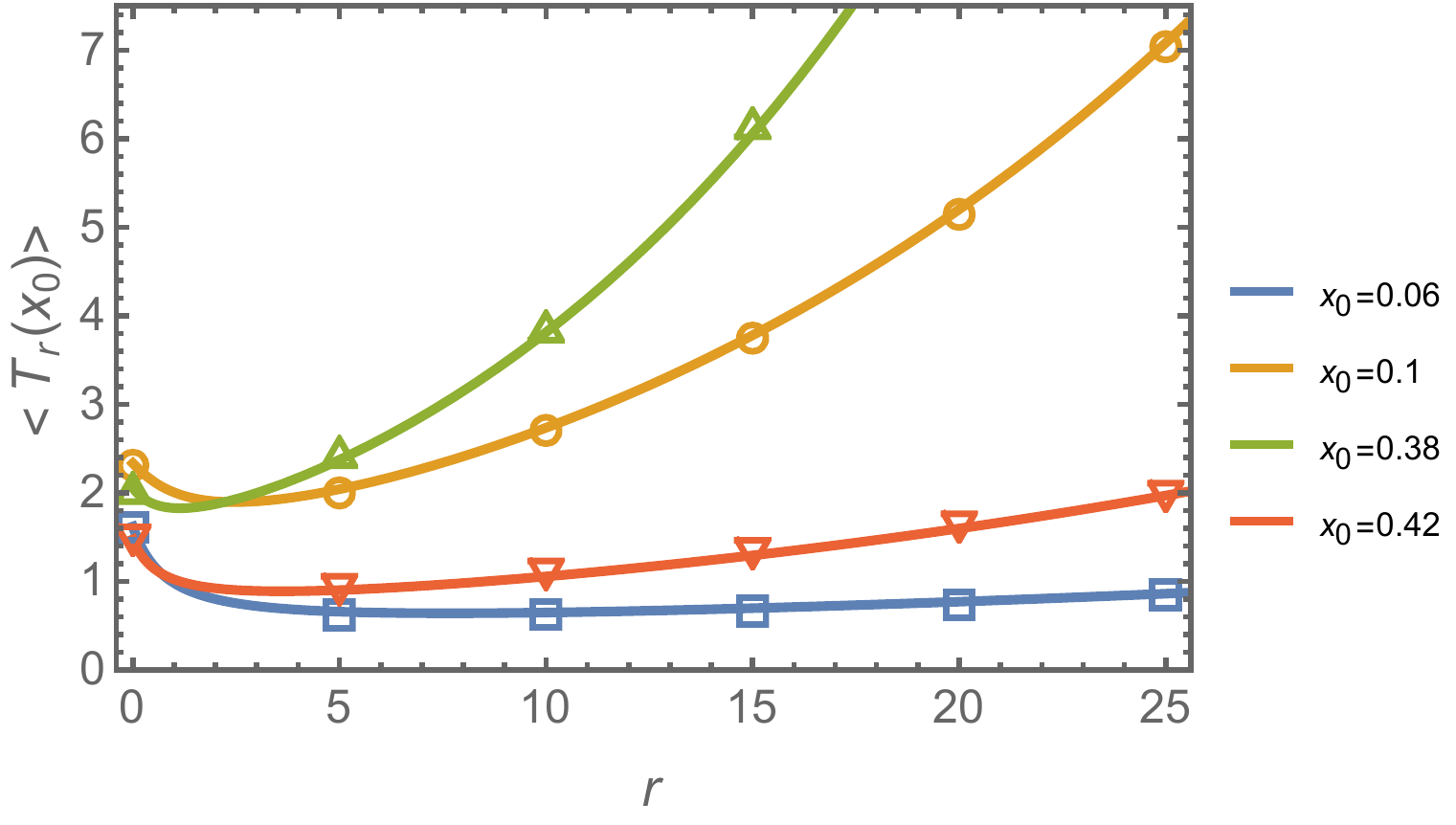}
		\includegraphics[width=0.45\textwidth]{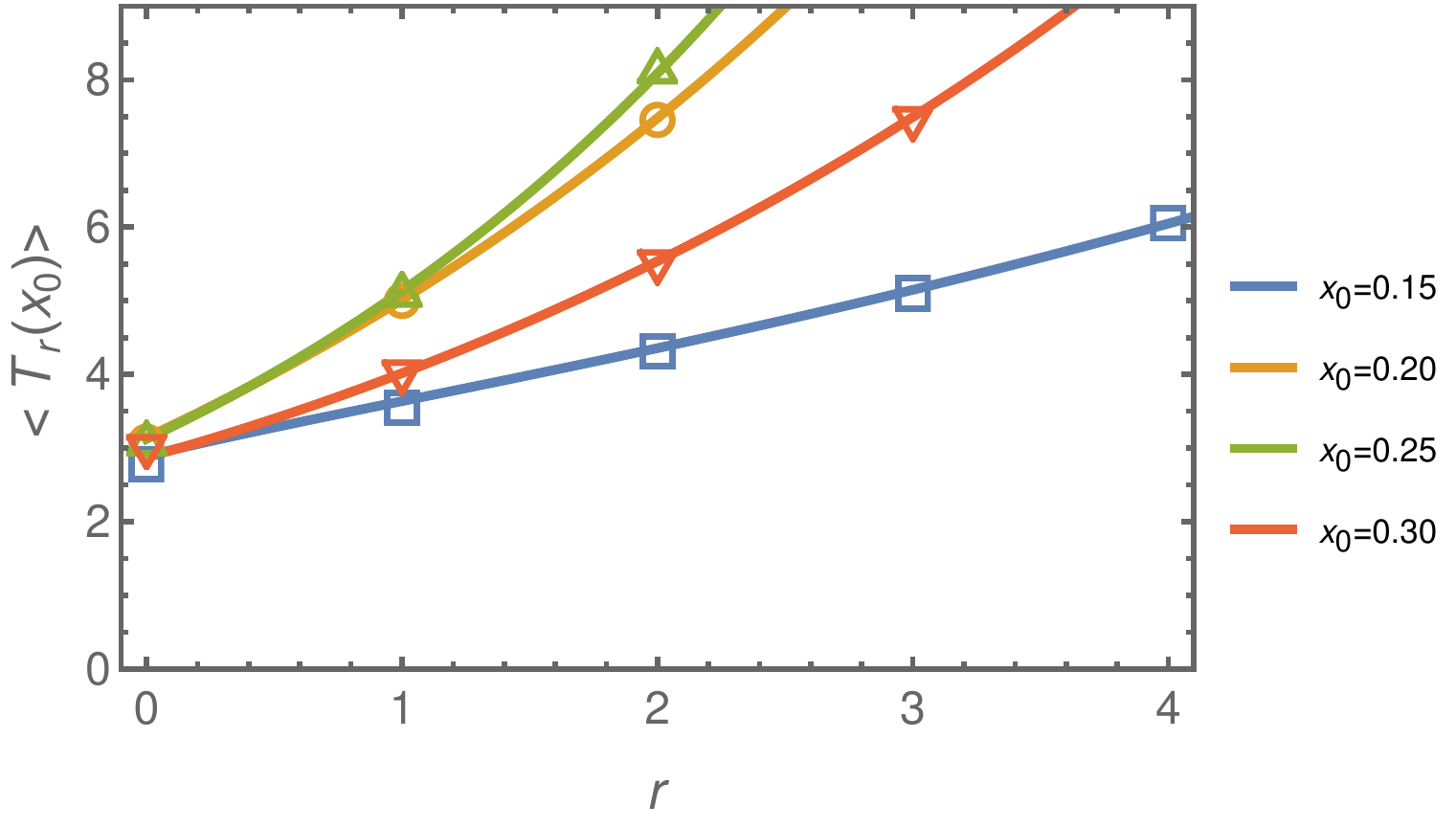}
	\caption{MFPTs predicted by Eq. \eqref{T2abs} (solid lines) are compared with the values obtained by  three-dimensional Brownian dynamics simulations (open symbols) as a function of the resetting rate for different starting positions, when resetting is useful (left) and when it is not (right). Absorbing-absorbing case with a expanding tube with $a=0.1$, $D_\lambda=0.01$, $L=0.5$, $\lambda$=0.2. The equality $CV=1$ for the reset-free process is achieved at $x_{0}$=0.1178 and $x_{0}$=0.3472.}
	\label{fig:reset_abs2}
\end{figure*}

We now turn our attention to the resetting induced transport process. As mentioned before, the renewal relations become crucial here. We start with the unconditional lifetime statistics of the particle inside the channel. In this case, the renewal relation is given by Eq. (\ref{firstpassage-renewal}) where we substitute expression for the underlying first passage time density from Eq. (\ref{eq:fsaa}) to find
\begin{widetext}
 \begin{align}
    \tilde{f}_{r}(s|x_0)=
    \frac{(s+r)[(\lambda L+a)\sinh(\beta x_{0})+a \sinh(\beta(L-x_{0})]}{s(\lambda x_{0}+a)\sinh(\beta L)+r[(\lambda L+a)\sinh(\beta x_{0})+a \sinh(\beta (L-x_{0})]},
\end{align}
where $\beta=\sqrt{(s+r)/D_\lambda}$. The mean first passage time under resetting is deduced by using the general relation (\ref{MFPTGen}), or $\langle T_{r}(x_{0}) \rangle=\lim_{s\rightarrow 0}s^{-1}[1-  \tilde{f}_{r}(s|x_0)]$, which gives
\begin{align}\label{T2abs}
    \langle T_{r}(x_{0}) \rangle=\frac{(\lambda x_{0}+a)\sinh(\alpha L)-a \sinh(\alpha (L-x_{0}))-(\lambda L+a)\sinh(\alpha x_{0})}{r[a \sinh(\alpha (L-x_{0}))+(\lambda L+a)\sinh(\alpha x_{0})]},
\end{align}
where $\alpha=\sqrt{r/D_\lambda}$.
\end{widetext}
In Fig. \ref{fig:pos_abs} (right), we have plotted $ \langle T_{r}(x_{0}) \rangle$ as a function of the initial/resetting coordinate, showing a larger lifetime near the center of the channel and a faster escape near the boundaries than in the corresponding case of the underlying process (Fig. \ref{fig:pos_abs} (left)). When we vary the resetting rate $r$, non-trivial effects can be observed. In  Fig. \ref{fig:reset_abs2} (left), resetting is seen to lower the mean first-passage time at small $r$, {\it i.e.}, it expedites the escape of the particle from the channel rendering a lower average lifetime. Nevertheless, in an other example shown in Fig. \ref{fig:reset_abs2}, $\langle T_{r}(x_{0}) \rangle$ increases monotonically with the rate $r$. While each variant above carries intriguing effects, it is nonetheless not evident how the parameters, {\it e.g.}, the initial position or the shape of the channel, set the criterion for resetting either to prolong or expedite the completion of the process. To characterize this transition, we do a detailed analysis of the resetting criterion in the next subsection.


\subsection{When is resetting useful?}
Resetting can often become beneficial to reduce the mean first passage time of a process, {\it i.e.}, $\langle T_r(x_0) \rangle < \langle T (x_0) \rangle$. Doing an expansion with respect to $r$ teaches us that resetting speeds up the mean first passage time when the coefficient of variation  $CV$ of the underlying process (with $r=0$) is larger than unity, where $CV$ is the ratio between the standard deviation of the first passage time, $\sigma(T)$, and its mean $\langle T\rangle$ \cite{PalReuveniPRL,inspection}. The criterion can also be recast as $\langle T^2(x_0)\rangle>2\langle T(x_0)\rangle^2$. From Eqs. (\ref{eq:Taa}) and (\ref{eq:T2aa}), this implies
\begin{align}\label{eq:critaa}
    (\lambda v+\tilde{a})
    \left[\lambda(1+v)(7-3v^2)+15\tilde{a}(1+v-v^2) \right]\geq \nonumber \\
    10v(1-v)
    \left[\lambda(1+v)+3\tilde{a} \right]^2,
\end{align}
where $\tilde{a}=a/L$ and $v=x_{0}/L$. One notices that the criterion is fulfilled in any channel as soon as the particle starts sufficiently close to one of the absorbing boundaries ($v\sim 0$ or $v\sim 1$) but not when it starts out in the middle. For starting positions near the center of the tube, increasing the reset rate increases the MFPT since any trajectory on both sides is going to take particle closer to the boundary and reset hinders it. But for starting positions closer to the boundaries, resetting decreases the lifetime because there are now many more possible trajectories that are taking the particle away from the boundary and resetting helps eliminate them. Setting $\lambda=0$, one recovers the case of the one-dimensional Brownian particle in an interval. In this case, Eq. (\ref{eq:critaa}) reduces to
\begin{equation}\label{eq:flat}
    5v^2-5v+1\geq 0,
\end{equation}
a relation which was previously derived in \cite{palprasad}. Putting in the parameters values of the system, we can find for which values of $x_{0}$ resetting is useful. For all the starting positions satisfying the criterion $CV>1$, there exists a finite optimal resetting rate (minimum of the curve) as seen in Fig. \ref{fig:reset_abs2} (left), while those with $CV<1$ have a minimum lifetime at $r=0$, as shown by Fig. \ref{fig:reset_abs2} (right).



\subsection{Optimal resetting rate}

\begin{figure*}[ht!]
\centering
		\includegraphics[width=0.45\textwidth]{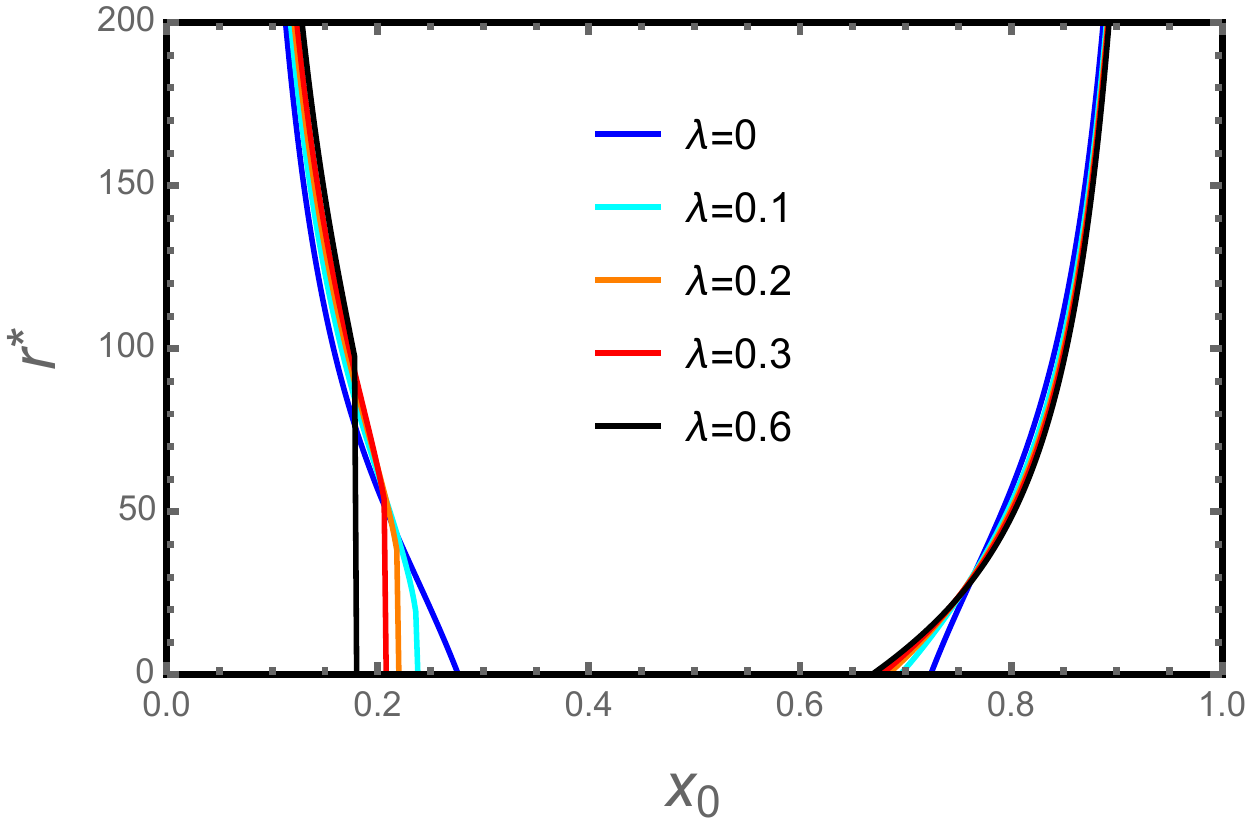}\hspace{0.3cm}
		\includegraphics[width=0.45\textwidth]{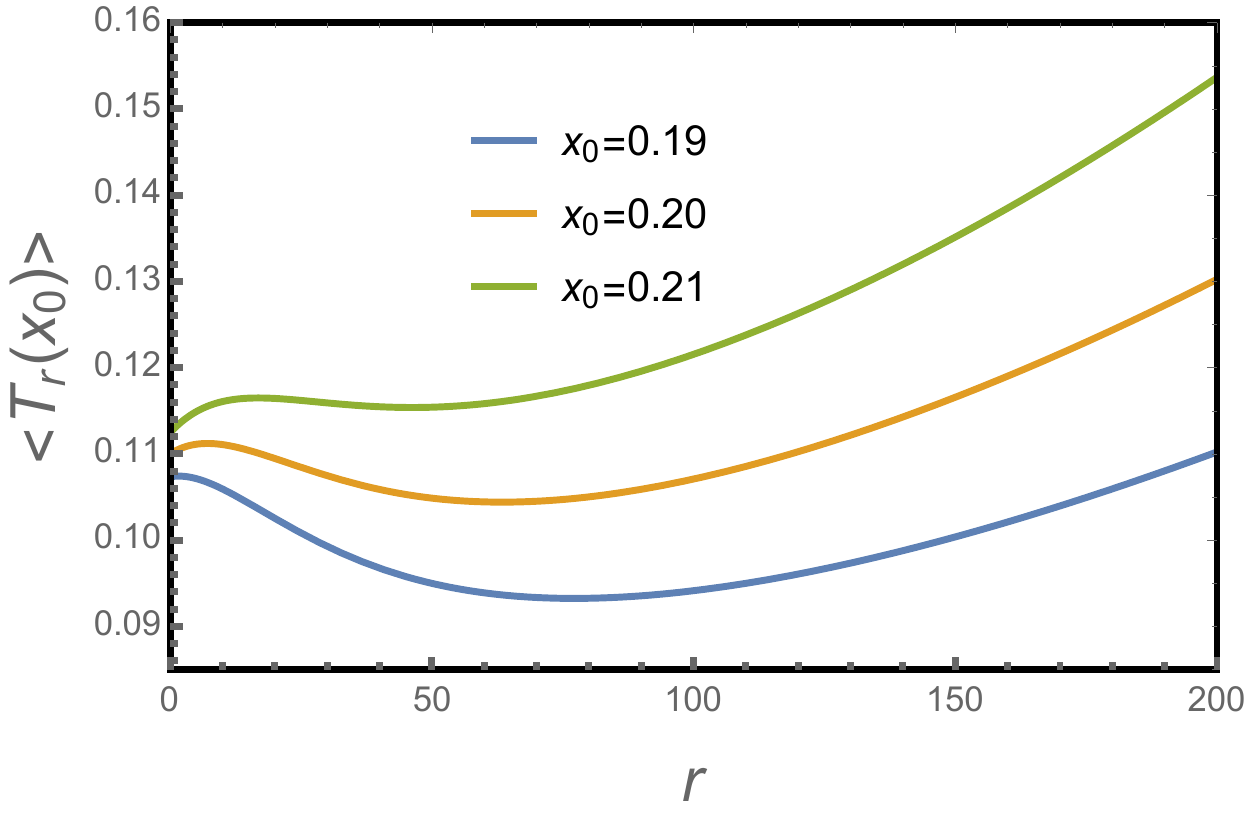}
	\caption{Left: Optimal restart rate $r^*$ vs. the starting position $x_0$, for various channel slopes $\lambda$ in the absorbing-absorbing case (the narrow boundary is on the left). The other parameters are $D_{\lambda}=1$, $a=0.1$, $L=1$. Right: Variations of the MFPT with $r$ when $x_0$ lies in the vicinity of the discontinuous transition, for $\lambda=0.3$.}
	\label{fig:roptabs2}
\end{figure*}

It is instructive to compute 
the optimal resetting rate, which is defined as the rate at which the MFPT reaches its minimum,
\begin{align}
    r^*=\underset{r\ge 0}{\arg \min} \langle T_r(x_0)\rangle,
\end{align}
the other parameters being fixed. The minimization of Eq. (\ref{T2abs}) can be carried out numerically and $r^*$ is displayed in Figure \ref{fig:roptabs2} (left) as a function of $x_0$, for several channel slopes $\lambda$. In the central part of the channel, $r^*=0$ as expected. As $x_0$ increases and approaches the wide boundary (to the right), a continuous transition to $r^*>0$ takes place at $x_0=x_0^c$. The latter position precisely corresponds to the point where $CV=1$. This transition holds for $\lambda=0$ (cylindrical channel) and $\lambda>0$. A noticeable effect of the slope is to extend the range of values of $x_0$ where resetting is beneficial, compared to the cylindrical geometry.

The situation is qualitatively different when $x_0$ decreases towards the narrow absorbing boundary (to the left). Interestingly, the optimal restart rate $r^*$ undergoes a {\it discontinuous} transition at a position $x_0^{c\prime}$ in this case, see Figure \ref{fig:roptabs2} (left). The origin of this abrupt jump, which is absent in cylinders but occurs as soon as $\lambda>0$, can be understood by representing the variations of $\langle T_r(x_0)\rangle$ with respect to $r$, as in the example of Figure \ref{fig:roptabs2} (right). For $x_0$ slightly above $x_0^{c\prime}=0.207...$, the optimal parameter is $r^*=0$ but the MFPT has developed a local, \lq\lq metastable" minimum at a finite rate $r_{local}$. As $x_0$ further decreases, the local minimum moves downward with respect to $\langle T_{r^*=0}(x_0)\rangle$ and becomes the absolute minimum after some point, where $\langle T_{r=0}(x_0^{c\prime})\rangle=\langle T_{r=r_{local}}(x_0^{c\prime})\rangle$. In contrast with the wide side, near the narrow boundary the extent of the region where restart is favoured shrinks as the slope $\lambda$ increases.

With a discontinuous transition, the criterion $CV=1$ cannot be used to obtain the exact value of $x_0^{c\prime}$. The condition $CV<1$ (a positive slope for the MFPT at $r=0$) is not sufficient to guarantee that $r^*=0$ in general, as a deeper minimum might exist for some other value of $r>0$. This is clearly the case for $\lambda=0.20$ in Figure \ref{fig:roptabs2} (right). Nevertheless, since the regime with two local minima exists for a tiny range of values of $x_0$, the criterion $CV=1$ provides a reasonable approximation for locating the discontinuous jump {\it a priori}. 

Discontinuous transitions for $r^*$ are relatively less common in resetting processes \cite{kusmierzprl,landaupal}. Analytical estimates can be made within a mean field theoretical treatment of resetting transition \cite{landaupal}. The joint observation of a continuous and a discontinuous transition is less common, but have actually been reported independently in different resetting problems related to the present one \cite{landaupal,ahmad}. Recently, diffusion in the presence of a roughly linear potential under resetting in an interval with absorbing-absorbing boundaries was studied in  \cite{ahmad}. If the potential is inclined, say, toward the boundary at $L$, the transition in $r^*$ as $x_0$ approaches this boundary is continuous, whereas it is discontinuous when $x_0$ approaches the other boundary, very much like in Figure \ref{fig:roptabs2} (left). That problem is qualitatively similar to ours, since, in our case, there actually exists an effective logarithmic potential created by the varying radius of the channel and which tends to push the particle toward the wide side located at $L$.

\begin{figure*}[ht!]
		\includegraphics[width=0.45\textwidth]{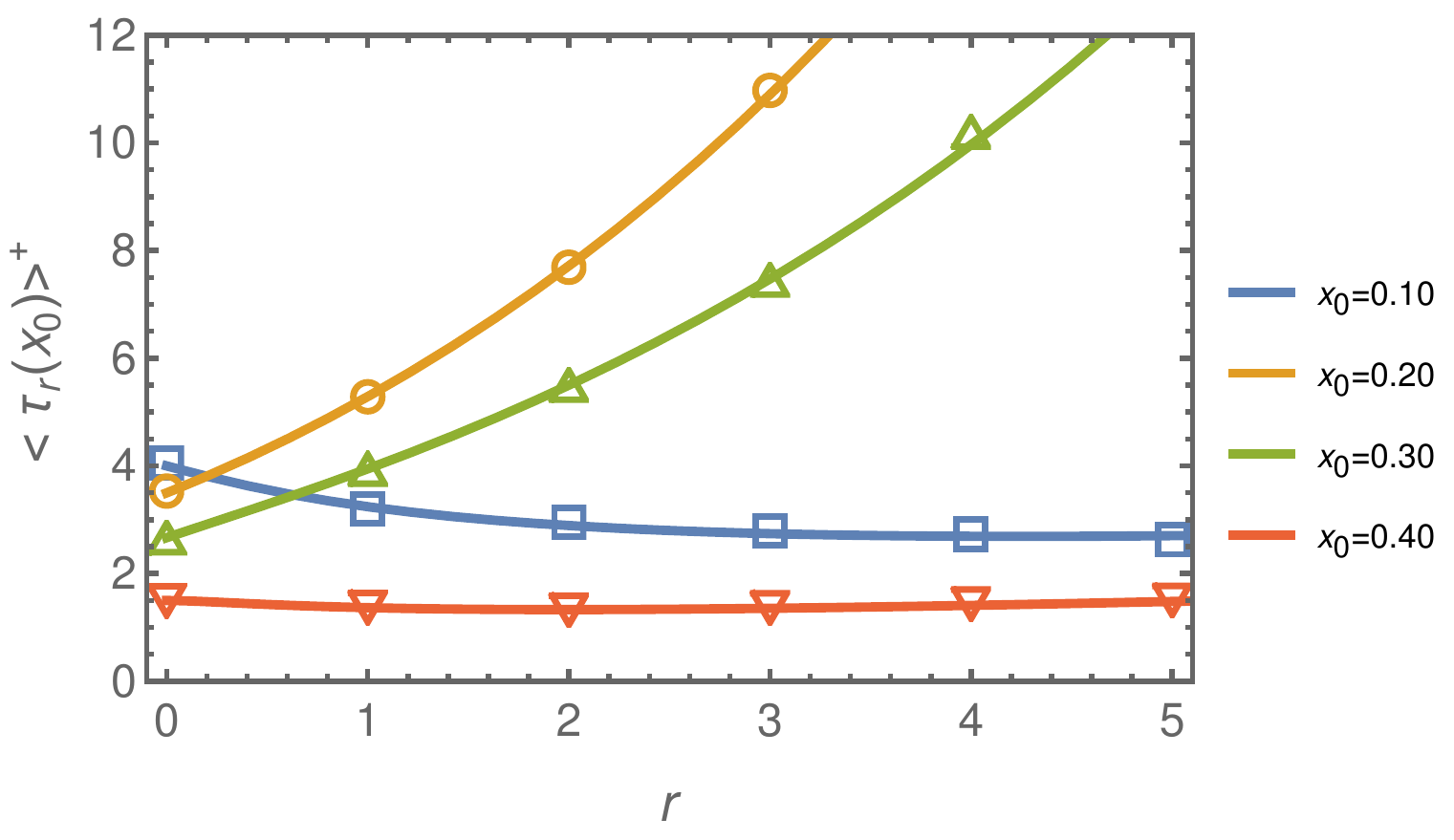}
		\includegraphics[width=0.45\textwidth]{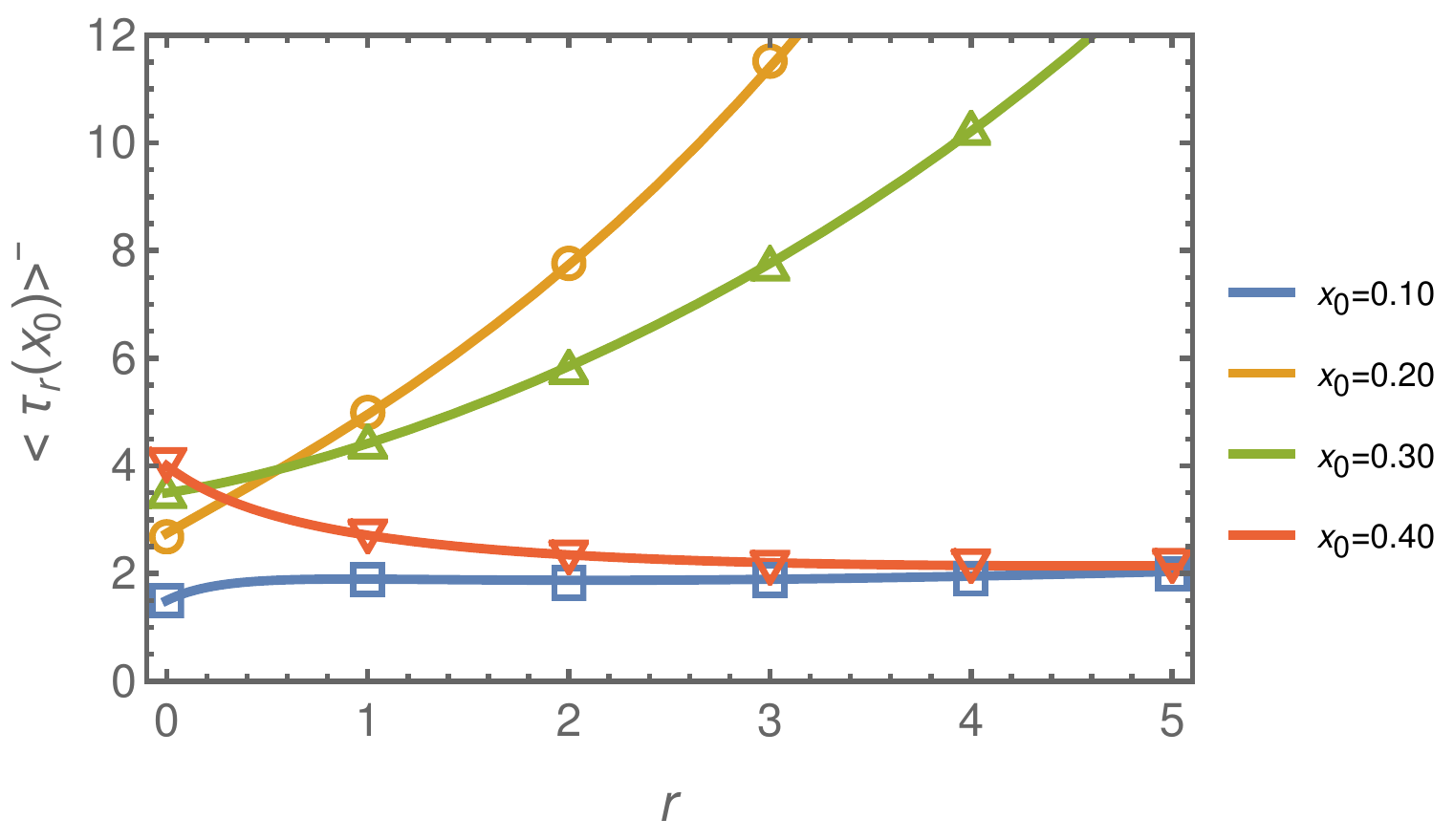}
	\caption{Conditional MFPT predicted by Eq. \eqref{ctpr} (left) and Eq. \eqref{ctmr} (right) as a function of the resetting rate for different starting positions:  absorbing-absorbing case for an expanding tube of parameters values $a=0.1$, $D_\lambda=0.01$, $L=0.5$, $\lambda$=0.2.}
	\label{fig:plusr}
\end{figure*}

\subsection{Conditional first passage or transition path times under resetting}
In this section, we study the conditional first passage time statistics and their behavior as a function of the resetting rate. Recalling the set of relations  (\ref{mean-exit-times-definition})-(\ref{conditionalFPT-renewal}) from Section IIB, and substituting the results for the unconditional and conditional first passage time densities for the underlying process, we obtain
\begin{widetext}
\begin{align}
    \tilde{f}^{+}_{r}(s|x_0)&=
    \frac{(s+r) \sinh(\beta x_{0})[(\lambda L+a)\sinh(\alpha x_0)+a\sinh(\alpha (L-x_0))]}{\sinh(\alpha x_0)[s(\lambda x_{0}+a)\sinh(\beta L)+r((\lambda L+a)\sinh(\beta x_{0})+a\sinh(\beta (L-x_0))]}, \\
    \tilde{f}^{-}_{r}(s|x_0)&=
    \frac{(s+r) \sinh(\beta (L-x_{0}))[(\lambda L+a)\sinh(\alpha x_0)+a\sinh(\alpha (L-x_0))]}{\sinh(\alpha (L-x_0))[s(\lambda x_{0}+a)\sinh(\beta L)+r((\lambda L+a)\sinh(\beta x_{0})+a\sinh(\beta (L-x_0))]},
\end{align}
with $\alpha=\sqrt{r/D_\lambda}$,  $\beta=\sqrt{(s+r)/D_\lambda}$, and where we have made use of Eqs. (\ref{conditionalFPT-renewal}), (\ref{eq:fsaa}) and (\ref{conditionalFPT-LT-underlying}). The conditional mean first passage times are directly obtained from there. Skipping details, we find
\begin{align}
       &\langle \tau_{r}(x_{0}) \rangle^{+}=\nonumber\\
       &\frac{-2(\lambda L+a)\sinh(\alpha x_0)+a\alpha L\cosh(\alpha (L-x_0))+2(\lambda x_0+a)\sinh(\alpha L)-2a\sinh(\alpha (L-x_0))-a\alpha x_{0} cosech(\alpha x_0)\sinh(\alpha L)}{2r[(\lambda L+a)\sinh(\alpha x_0)+a\sinh(\alpha(L-x_0))]},
\label{ctpr}       
\end{align}
and
\begin{align}
    &\langle \tau_{r}(x_{0}) \rangle^{-}=\nonumber\\
    &\frac{-2a \sinh(\alpha (L-x_0))+(\lambda L+a)\alpha x_0 \cosh(\alpha x_0)+2(\lambda x_0+a)\sinh(\alpha L)-(\lambda L+a)(2+\alpha (L-x_0) \coth(\alpha (L-x_0)))\sinh(\alpha x_0)}{2r[a\sinh(\alpha (L-x_0))+(\lambda L+a)\sinh(\alpha x_0)]}.
\label{ctmr}
\end{align}
\end{widetext}

As expected, these expressions reduce to Eqs. (\ref{ctm}) in the $r \to 0$ limiting case. The conditional times are plotted as a function of the resetting rate in Fig. \ref{fig:plusr}, showing an excellent agreement with simulations.

\begin{figure*}[ht!]
		\includegraphics[width=0.45\textwidth]{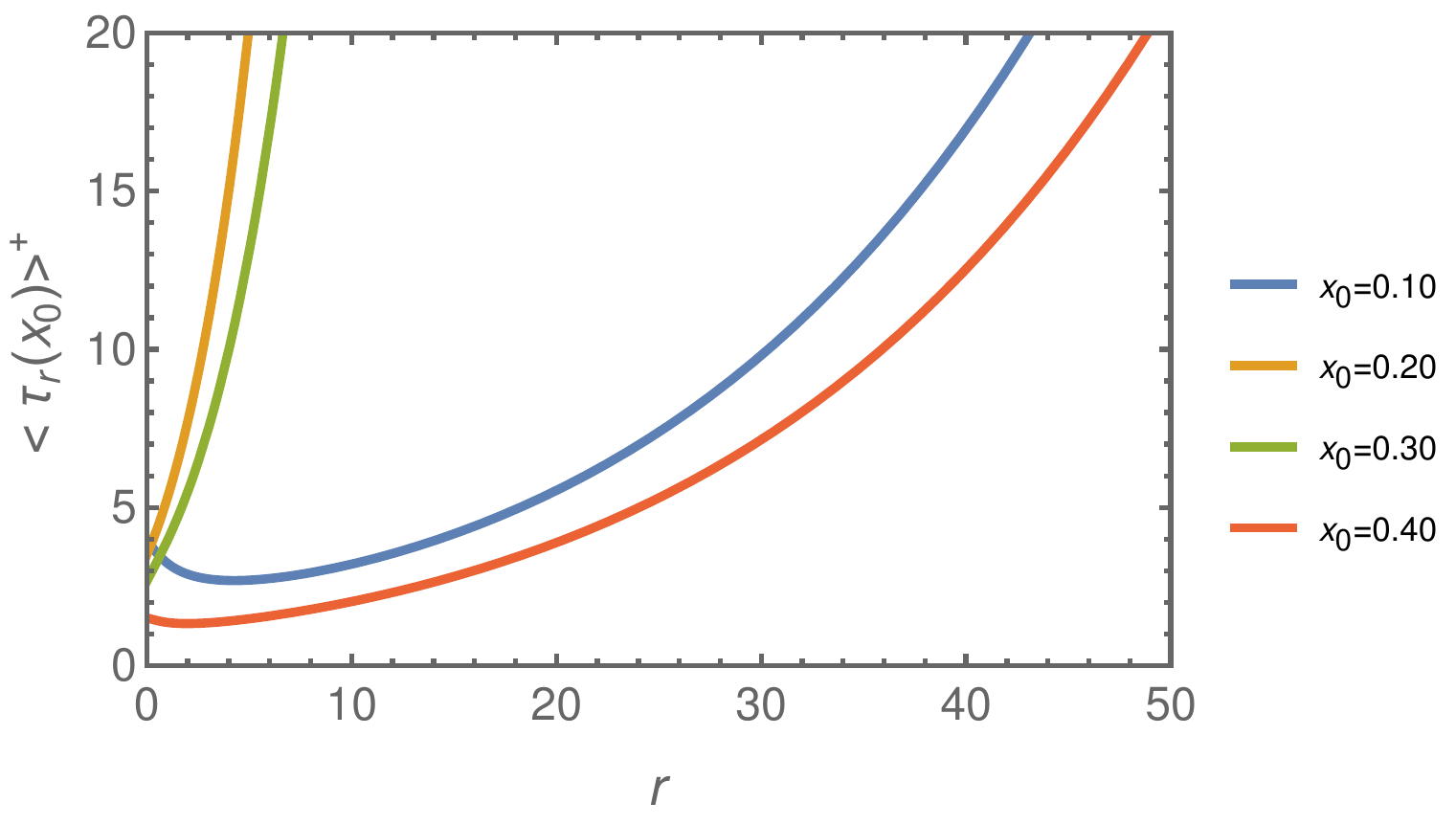}
		\includegraphics[width=0.45\textwidth]{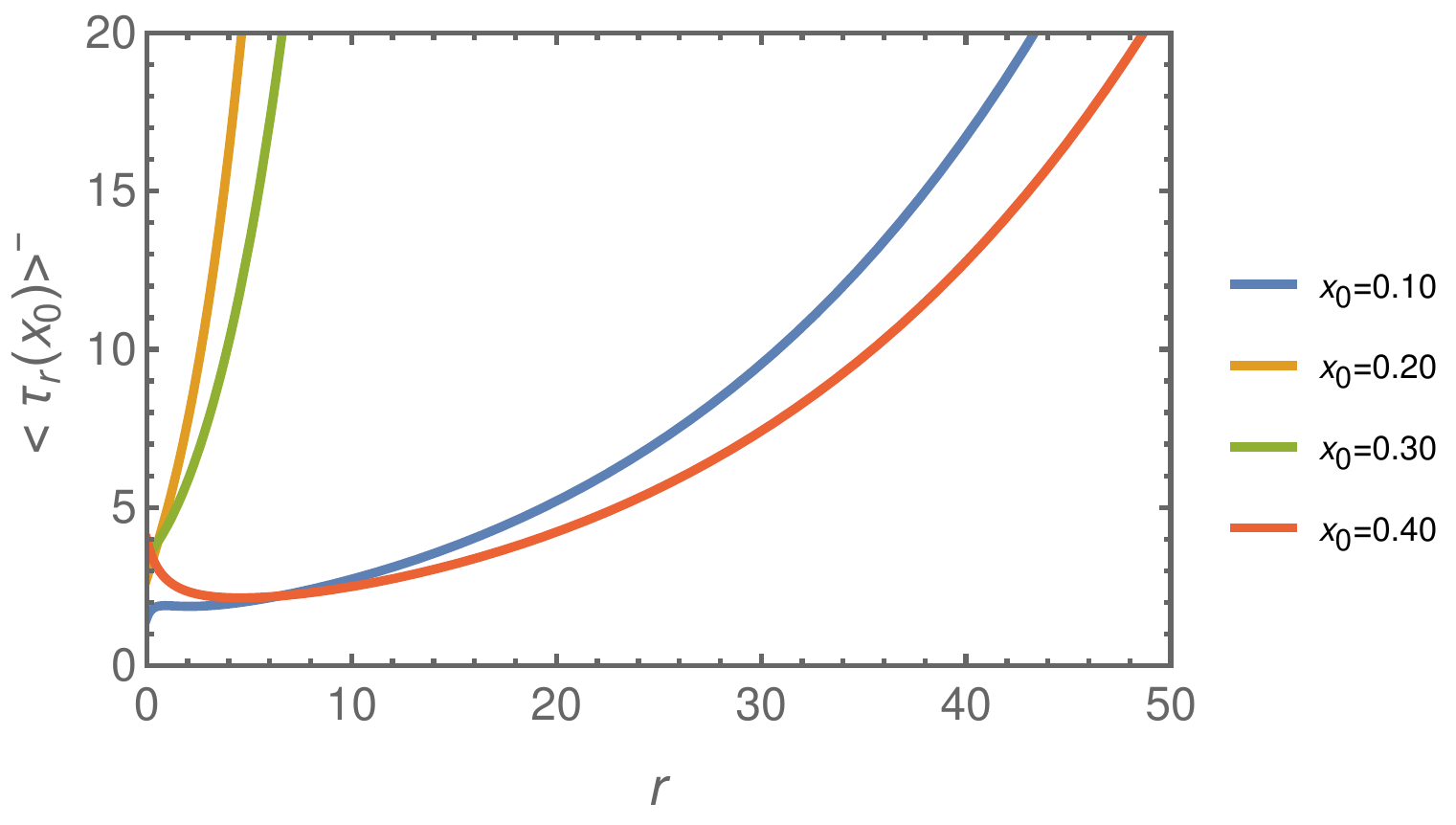}
	\caption{Same as Figure \ref{fig:plusr}, over a wider interval of values of $r$ capturing both monotonic and non-monotonic behavior.}
	\label{fig:plusr-0}
\end{figure*}

Similar to the unconditional mean-first passage time, the conditional times also display both monotonic and non-monotonic behavior as a function of the resetting rate  (see Fig. \ref{fig:plusr-0} for a wider range of the parameter $r$). The introduction of resetting can either increase or decrease the conditional times at small $r$, as shown in Fig. \ref{fig:plusr}, however for large $r$, they always increase. 
Furthermore, the conditional mean times can also be optimized with the introduction of resetting in some cases that are quite counter-intuitive, {\it e.g.}, when particle starts and resets at the opposite end of the desired escape side. See for instance the curve corresponding to $x_0=0.1$ in Fig. \ref{fig:plusr} (left). Even though the particle starts very close to the left boundary, resetting can expedite its escape from the right boundary. Similarly, take a look at the plot corresponding to $x_0=0.4$ in Fig. \ref{fig:plusr} (right). Resetting can expedite the escape from the left boundary even for those trajectories which start and reset at a point very close to the right boundary.

This non-intuitive behavior can be explained qualitatively in the following way. Statistically, all the trajectories set out randomly to find the targets. Some of these realizations spend more time going to the opposite direction of the desired boundary (where we expect them to be absorbed) without being trapped by the other absorbing end meanwhile. Naturally, such trajectories render huge fluctuations in the first passage time. In addition, expansion of the tube also creates more space for exploration, contributing to these fluctuations as well. In such cases, resetting even to a close point to the non-desired end can help 
to mitigate such large fluctuations (and avoid looping or transient excursions) -- a signature characteristic also can be seen for the unconditional time. Eventually, the particles translocate through the tube and finally reach the desired end after a considerable time. While this is not a typical behavior, nonetheless it happens strongly induced by resetting. This counter-intuitive observation, which does not arise in the reset-free process, is one noteworthy result of our work.

\section{First passage observables in a three dimensional Cone with One absorbing and One reflective boundary}\label{sec:reflabs}

\begin{figure*}[ht!]
\centering
		\includegraphics[width=0.38\textwidth]{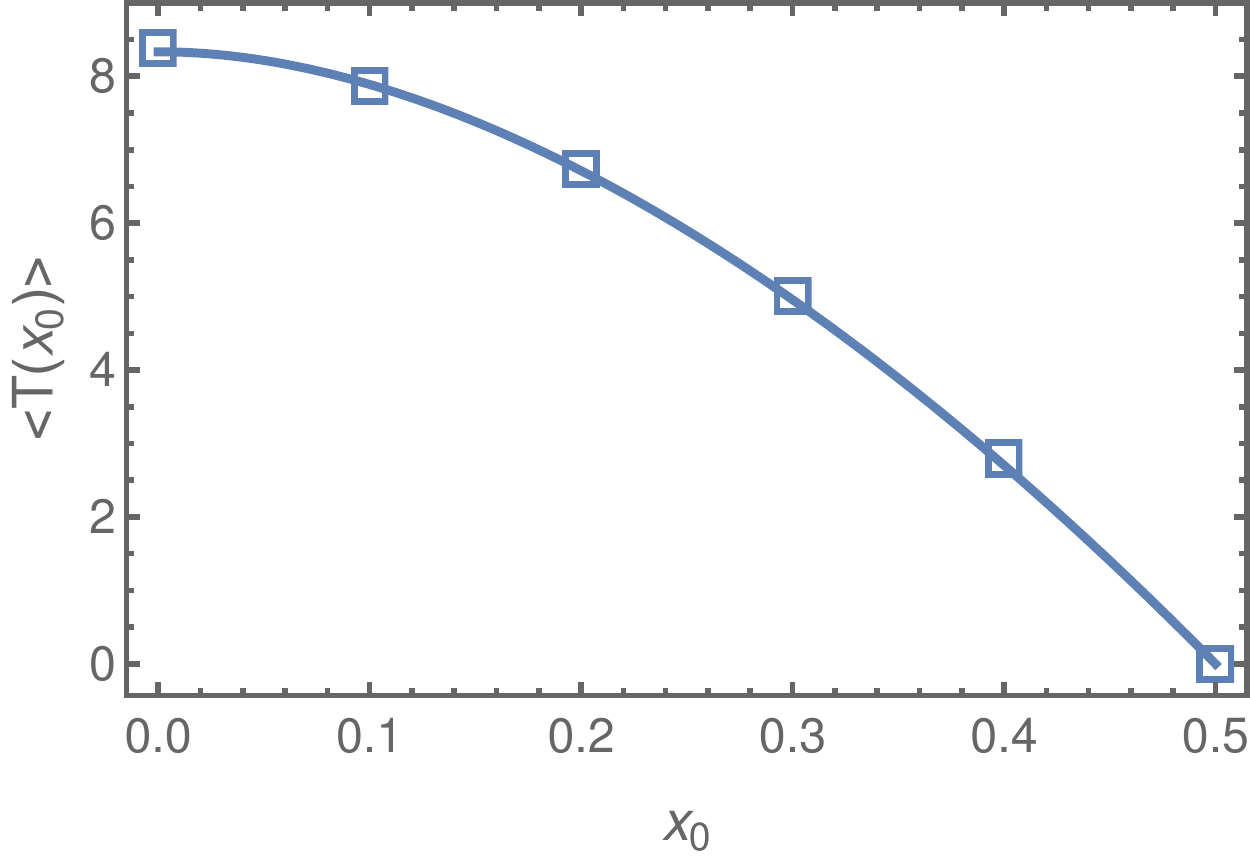}
		\includegraphics[width=0.45\textwidth]{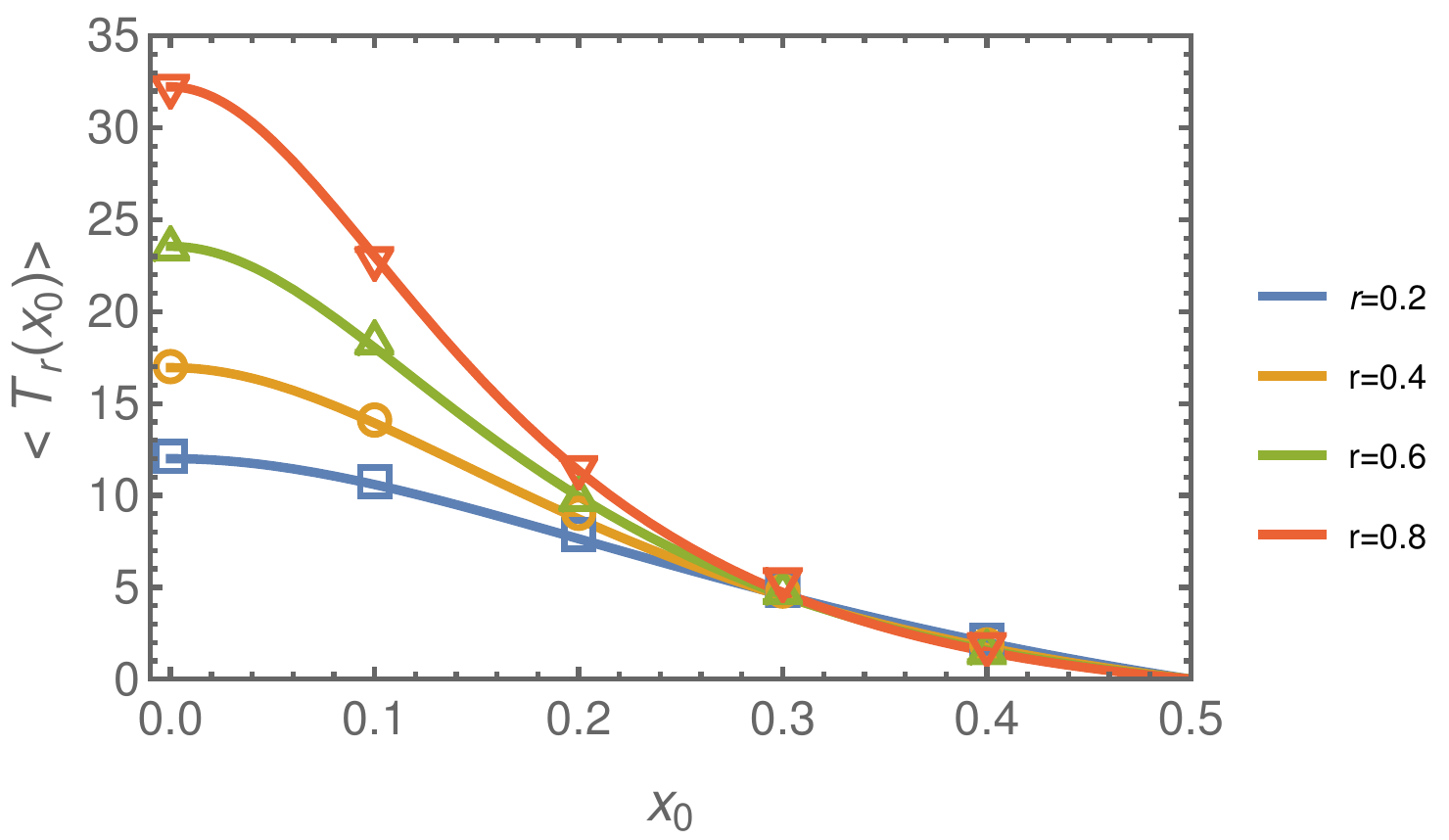}
	\caption{The MFPT predicted by Eqs. \eqref{eq:T1ar} and (\ref{MFPT-r-reflect-abs}) for expanding tubes (continuous lines) are compared with the values obtained by  three-dimensional Brownian dynamics simulations (open symbols) as function of the starting position, without resetting (left) and with resetting (right), for the reflecting-absorbing case. The parameters are $a=0.1$, $D_\lambda=0.01$, $L=0.5$, $\lambda$=0.2.}
	\label{fig:ref1}
\end{figure*}

In this section, we consider a 3D narrow-to-wide expanding cone which has a reflective boundary at one end ($x=0$) and
an absorbing boundary at the other ($x=L$). Later, we will also consider the wide-to-narrow geometry, a case which is physically different since the absorbing boundary switches from the wide side to the narrow side. For direct comparison, it is convenient to set the parameter $a$ that denotes the narrow diameter in both geometries. The wide-to-narrow quantities can therefore be deduced by making the transformations $\lambda\rightarrow-\lambda$ and $a\rightarrow \lambda L+a$ into the narrow-to-wide expressions (with $\lambda\ge0$ in all cases), as discussed in Appendix \ref{appendixa}. 

The particle starts from $x_0$, but it can escape only from one side and thus the conditional times are identical to the unconditional times. The absorbing boundary condition remains the same {\it i.e.}, $c(L,t)=0$, while the reflecting boundary condition imposes the probability current to vanish at  $x=0$ such that $J(0,t)=0$ where
$
    J(x,t)=-D_\lambda R^{2}(x)\frac{\partial}{\partial x}\left[\frac{{c}(x,t)}{R^{2}(x)} \right]
$. 
Recalling $c(x,t)=R(x)g(x,t)$ from Eq. (\ref{sep-var}), the reflecting boundary condition simplifies to
\begin{align}
 \frac{\partial g}{\partial x}\bigg|_{x=0}=(\lambda g/a )\bigg|_{x=0}.
\end{align}
As before, the next step is to compute the concentration $c(x,t)$ for which we need to evaluate the unknown constants of the solutions (\ref{eq:sect}) and (\ref{eq:sect1}). To this end, we use the Laplace transform of the above-mentioned boundary conditions and obtain
\begin{gather}
    c_{1}=\frac{-(\lambda-ua)\sinh(u(L-x_{0}))}{2uD_\lambda(\lambda x_{0}+a)[\lambda \sinh(uL)+ua \cosh(uL)]}, \\
    c_{2}=\frac{(\lambda+ua)\sinh(u(L-x_{0}))}{2uD_\lambda(\lambda x_{0}+a)[\lambda \sinh(uL)+ua \cosh(uL)]}, \\
    c_{3}=\frac{e^{uL}[\lambda \sinh(ux_{0})+ua \cosh(ux_{0})]}{2uD_\lambda(\lambda x_{0}+a)[\lambda \sinh(uL)+ua \cosh(uL)]}, \\
     c_{4}=\frac{-e^{-uL}[\lambda \sinh(ux_{0})+ua \cosh(ux_{0})]}{2uD_\lambda(\lambda x_{0}+a)[\lambda \sinh(uL)+ua \cosh(uL)]},
\end{gather}
where $u=\sqrt{s/D_\lambda}$. It is worth noting that this result allows us to take any value of $x_0$ within the channel, which generalizes the one obtained in \cite{BDB2017} where the result is restricted to $x_0=0$.

\subsection{Current and first passage times of underlying process}
Since the particle can escape only through the right boundary $L$, the total flux is given from using Eq. (\ref{right-J-r=0}),
\begin{gather}
    j(s|x_0)
    =\frac{(\lambda L+a)}{(\lambda x_{0}+a)}\frac{[\lambda \sinh(ux_{0})+ua \cosh(ux_{0})]}{[\lambda \sinh(uL)+ua \cosh(uL)]},
\end{gather}
which is equal to the first passage time density in Laplace space.
The first and the second moment of the underlying process are then easy to compute by expanding the above expression in powers of $u^2$. One obtains,
\begin{align}\label{eq:T1ar}
    \langle T(x_{0})\rangle=\frac{L^2(\lambda L+3a)}{6D_\lambda(\lambda L+a)}-\frac{x_0^2(\lambda x_0+3a)}{6D_\lambda(\lambda x_0+a)},
\end{align}
and 
\begin{align}\label{eq:T2ar}
   \langle T^2(x_{0})\rangle&=\frac{x_0^4(\lambda x_0+5a)}{60D^2(\lambda x_0+a)}-\frac{L^4(\lambda L+5a)}{60D^2(\lambda L+a)} \nonumber\\
    &+\frac{L^4(\lambda L+3a)^2}{18D^2(\lambda L+a)^2}-\frac{L^2x_0^2(\lambda L+3a)(\lambda x_0+3a)}{18D^2(\lambda L+a)(\lambda x_0+a)}.
\end{align}
When $x_{0}$=0, the first passage density and MFPT are in agreement with previously known results \cite{BDB2017}:
\begin{align}
    \tilde{f}(s|x_{0}=0)&=\frac{u(\lambda L+a)}{\lambda \sinh(uL)+ua\cosh(uL)}, \\
    \langle T(x_{0}=0)\rangle&=\frac{L^{2}(3a+\lambda L)}{6D_\lambda(a+\lambda L)}.
\end{align}
In Fig. \ref{fig:ref1} (left), we compare our analytical prediction for the first moment of the first-passage time, given by Eq. (\ref{eq:T1ar}) with the results obtained from three-dimensional Brownian dynamics simulations. Again, one can see excellent agreement between the theoretical predictions and simulation results.   


\subsection{First passage time statistics under resetting}
\begin{figure}[h!]
\centering
	
		\includegraphics[width=8.7cm]{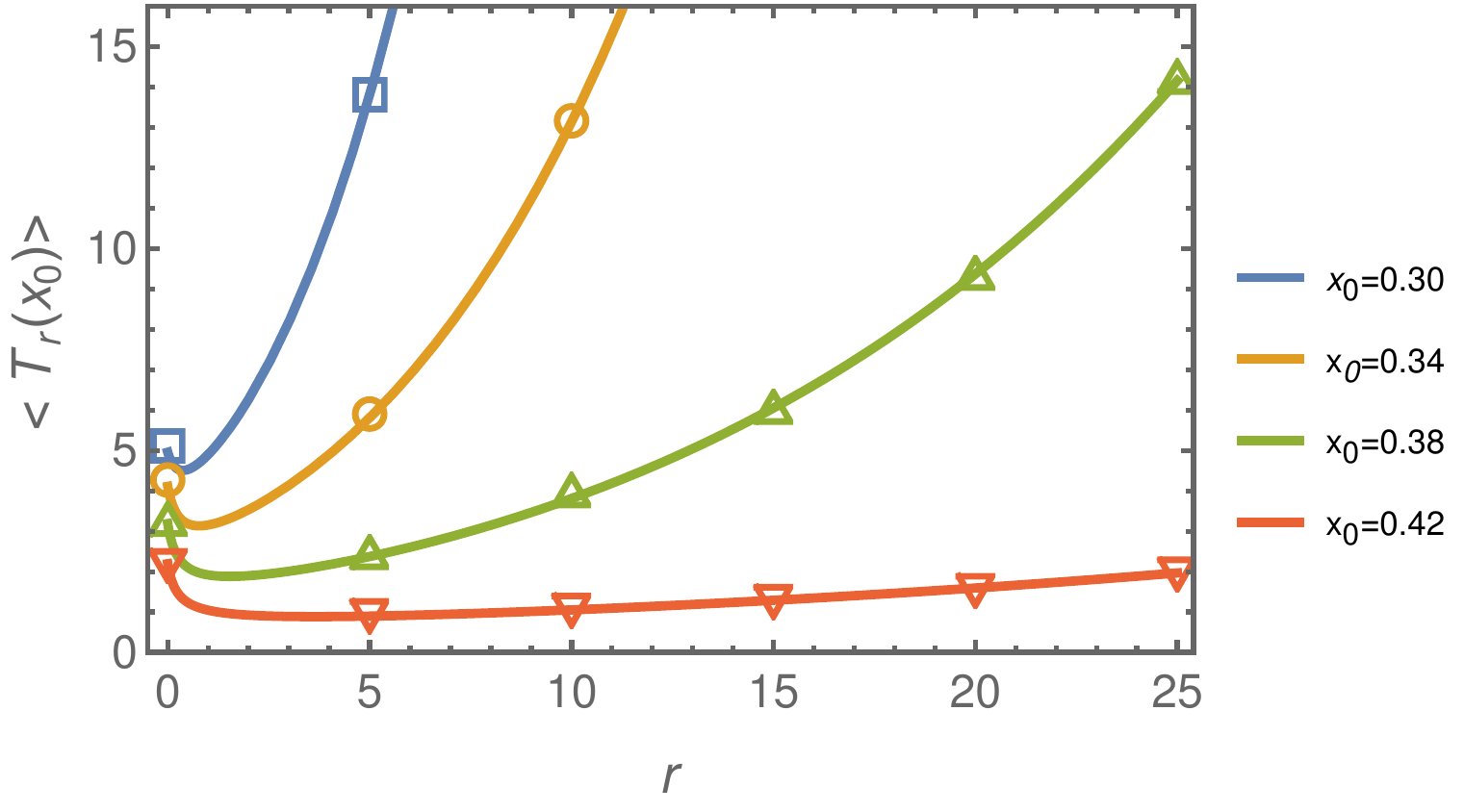}
		\includegraphics[width=8.7cm]{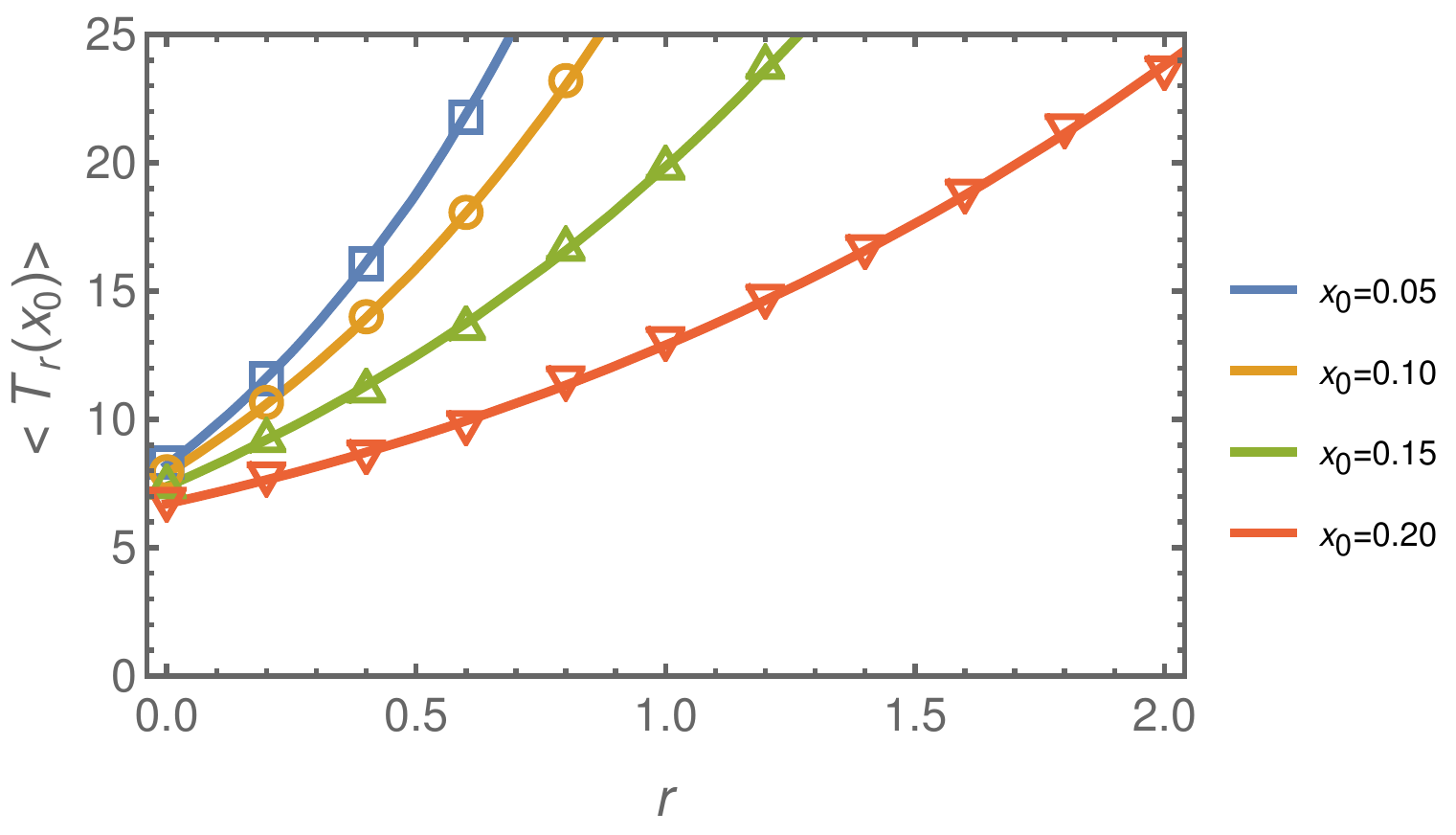}

	\caption{MFPT predicted by Eq. \eqref{MFPT-r-reflect-abs} (continuous lines) is compared with the values obtained by three-dimensional Brownian dynamics simulations (open symbols) as function of reset rate for starting positions when resetting is useful (top) vs. when resetting is not useful (bottom): reflecting-absorbing case with expanding tube for parameter values a=0,1, $D_\lambda=0.01$, L=0.5, $\lambda$=0.2. }
	\label{fig:ref2}
\end{figure}

To obtain the exit time statistics under resetting, we again use the renewal relations obtained in Section \ref{FJ-reset-renewal}.
Here, the first passage time density and the mean escape time have the following forms, respectively:
\begin{widetext}
\begin{gather}
    \tilde{f}_{r}(s|x_0)= \frac{(s+r)(\lambda L+a)[\lambda \sinh(\beta x_{0})+\beta a \cosh(\beta x_{0})]}{r(\lambda L+a)[\lambda \sinh(\beta x_{0})+\beta a \cosh(\beta x_{0})]+s(\lambda x_{0}+a)[\lambda \sinh(\beta L)+\beta a \cosh(\beta L)]},
\end{gather}
and 
\begin{align}
    \langle T_{r}(x_{0})\rangle=\frac{(\lambda x_{0}+a)[\lambda \sinh(\alpha L)+\alpha a \cosh(\alpha L)]-(\lambda L+a)[\lambda \sinh(\alpha x_{0})+\alpha a \cosh(\alpha x_{0})]}{r(\lambda L+a)[\lambda \sinh(\alpha x_{0})+\alpha a \cosh(\alpha x_{0})]},
    \label{MFPT-r-reflect-abs}
\end{align}
\end{widetext}
where $\beta=\sqrt{(s+r)/D_\lambda}$ and $\alpha=\sqrt{r/D_\lambda}$. We have verified the theoretical expression for the mean escape time  given by Eq. (\ref{MFPT-r-reflect-abs}) against numerical simulation in Fig. \ref{fig:ref1} right, finding an excellent agreement. In Fig. \ref{fig:ref2}, we represent $\langle T_r \rangle$ as a function of the resetting rate. As in the two-absorbing boundary case, resetting can lower the mean escape time if the particle starts close enough to the absorbing boundary (top panel), while from far away positions (bottom panel), resetting only prolongs the escape time. This transition can again be understood by looking at the $CV>1$ criterion, which we study next.

\subsection{When is resetting useful?}
As mentioned before, the introduction of resetting is able to reduce the mean escape time if the relative fluctuations ($CV$) of the escape time for the underlying process are greater than unity. Recasting the criterion into the form $\langle T^2(x_0)\rangle>2\langle T(x_0)\rangle^2$ and using Eqs. (\ref{eq:T1ar}) and (\ref{eq:T2ar}), we find
\begin{align}\label{eq:T3ar}
    \frac{6}{5}v^4\left(\frac{\lambda v+5\tilde{a}}{\lambda v+\tilde{a}}\right)&\geq
        4\left[v^2\frac{\lambda v+3\tilde{a}}{\lambda v+\tilde{a}}-\frac{\lambda +3\tilde{a}}{2(\lambda+\tilde{a})}\right]^2
        \nonumber \\ 
&+\frac{\lambda^2+6\tilde{a}\lambda-15\tilde{a}^2}{5(\lambda+\tilde{a})^2},
\end{align}
with $v=x_0/L$. For $\lambda=0$ , this reduces to: $5v^{4}-6v^{2}+1 \leq 0$. For $v=0$, the criterion (\ref{eq:T3ar}) is never satisfied as $\langle T^2(x_0=0)\rangle-2\langle T(x_0=0)\rangle^2=-\frac{L^4}{60D_\lambda^2}\frac{\lambda+5\tilde{a}}{\lambda +\tilde{a}}<0$. In this case, since
the initial/resetting location is at the origin, where the reflecting boundary is placed, if one resets to this boundary, reset cannot expedite the completion as every trajectory starting here is a favourable trajectory for the particle to reach the absorbing boundary. 
There exists a finite optimal resetting rate for all starting positions satisfying the criterion, as shown in Fig. \ref{fig:ref2} (top panel).

\subsection{Optimal resetting rate}
\begin{figure}[t!]
\centering
		\includegraphics[width=0.45\textwidth]{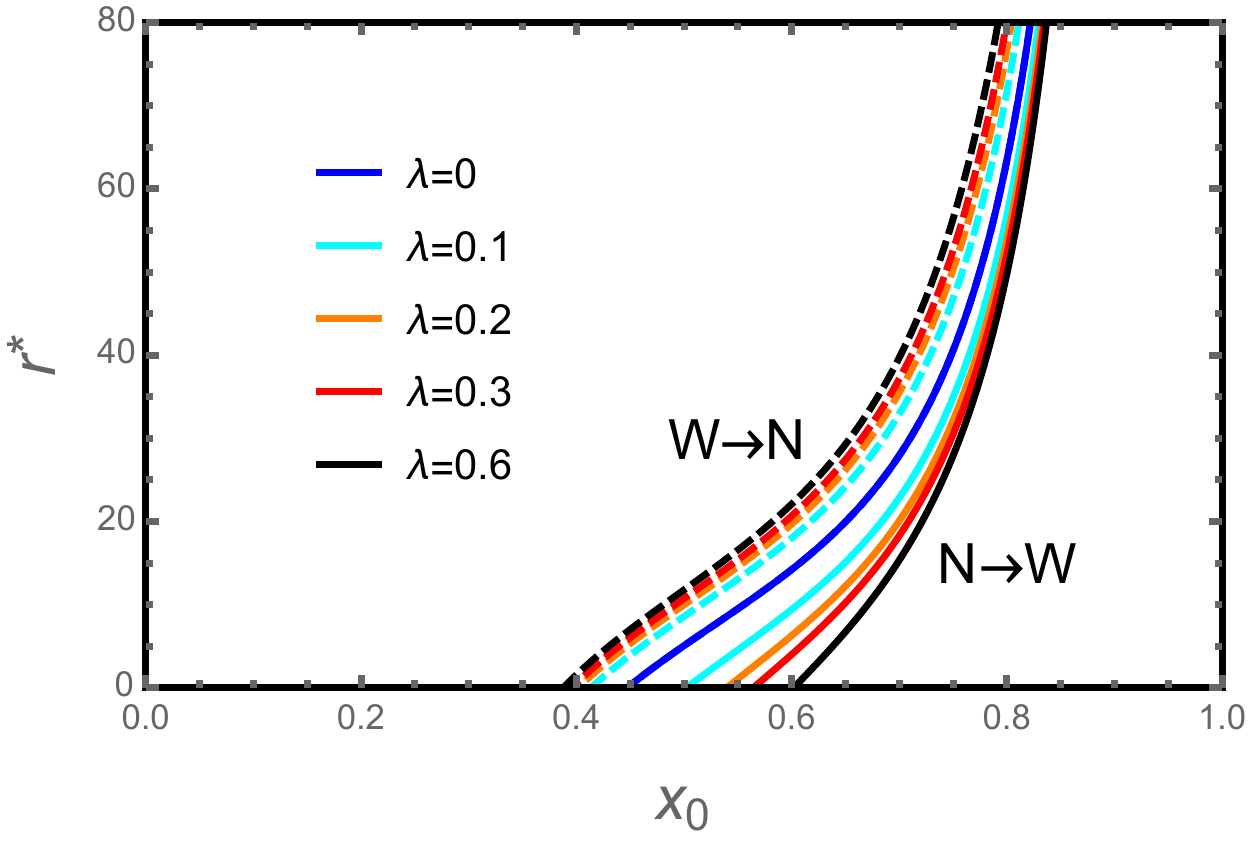}
	\caption{ Optimal restart rate $r^*$ vs. the starting position $x_0$, for various channel slopes $\lambda$ in the absorbing-reflective case. The absorbing boundary is on the right ($x_0=1$). The solid lines correspond to the N$\rightarrow$W geometry and the dashed ones to the W$\rightarrow$N case. $D_{\lambda}=1$, $L=1$, and the narrow diameter is $a=0.1$.} \label{fig:roptabs1}
\end{figure}

Variation of the optimal resetting rate with the initial position is displayed in Figure \ref{fig:roptabs1}, where the effects of the channel geometry can also be appreciated. At $\lambda=0$, one recovers a continuous transition to non-vanishing values of $r^*$, which occurs at
$x_0^c/L=1/\sqrt{5}=0.4472...$. A well-known result for the 1D interval with a single absorbing boundary \cite{ahmad,freezing}.

In the narrow-to-wide geometry (solid lines), a finite $\lambda$ results in a shift of $x_0^c$ toward the absorbing boundary. This can be understood by the fact that, once again, the expanding channel generates an effective potential which biases freely diffusing Brownian particles toward the absorbing boundary. Owing to such attraction, one expects lower fluctuations of the first passage time. The larger the channel slope, the stronger the effective potential, and resetting becomes less needed. This finding agrees qualitatively with the picture that emerges in the study of diffusion under resetting on the semi-infinite line with a constant drift toward a target \cite{ray}: in that problem, as the drift velocity increases at fixed $x_0$, the optimal resetting rate decreases, until it vanishes at a critical velocity. Similar findings were reported by using other attractive potentials \cite{ahmad2}.

Conversely, in the wide-to-narrow configuration (dashed lines of Fig. \ref{fig:roptabs1}), the effective potential pushes the particle in the direction opposite to the absorbing boundary. In this case, restart is beneficial over a larger range of values of $x_0$. It is worth mentioning that the transition remains continuous: no discontinuous jumps occur, unlike in Figure \ref{fig:roptabs2} (left). This highlights the fact that the number of absorbing boundaries has quite non-trivial consequences on the optimization of resetting processes. 
Finally, as $x_0\rightarrow0$, all the cases converge toward the optimal rate for a $1d$ interval ($\lambda=0$): when the particle starts very close to the target boundary, the channel geometry becomes irrelevant in the optimization process. This feature is also shared by the absorbing-absorbing case (see Fig. \ref{fig:roptabs2}-left).
\section{Conclusions}
\label{conclusions}
We have studied the transport of diffusing particles between the left and right ends of a 3D expanding or narrowing conical tube/channel. Such transitions are ubiquitous in nature and have important applications in various biochemical processes such as isomerization reactions, enzymatic catalysis, barrier crossing or channel facilitated transport of solutes. In addition to the diffusion, the particles also experience an intermittent transient dynamics, namely, resetting which puts them back to some preferred locations. Usually, in the reset-free process, a particle can escape from both boundaries due to the entropic potential which can also render a preference in the escape direction. In the resetting process, wandering trajectories causing large fluctuations in the escape time are eliminated, causing a significant reduction in the overall escape time.

We approached the problem using an effective one-dimensional description of the particle diffusion along the tube axis in terms of the modified Fick-Jacobs equation. There, the three dimensional effects are replaced by the so-called entropic potential that essentially plays a crucial role in the particle dynamics. As such, the diffusion now occurs with an effective diffusivity entering into the modified Fick-Jacobs equation, which is smaller than the particle diffusivity in a cylindrical tube. This is one way how the varying tube geometry manifests itself in the direct transits. The presence of resetting creates an influx of probabilities to the resetting coordinate by withdrawing probabilities from the rest of the free space -- this necessitates further modification of the 
Fick-Jacobs formalism for the resetting process. Using this, we presented a comprehensive study of escape properties. In particular, we (re)derived renewal relations for the (un)conditional escape times which become crucial in understanding the behavior of such observables in the presence of resetting. In particular, we computed the average lifetime that the particle spends inside the conical channel before it exits through one of the end points. 

We find that the resetting can expedite the crossing/escape time of the particle -- a hallmark feature that is shared by many other stochastic processes subject to resetting. This feature, however, is not arbitrary and it has been repeatedly demonstrated that resetting works in favour when stochastic fluctuations in the completion time of a random process are large, {\it i.e.}, $CV>1$. We explicitly verify this criterion for both the expanding and narrowing conical tubes. The effects of resetting are further analyzed by observing the behavior of optimal resetting rate $r^*$ that minimizes the mean escape time. Interestingly, the optimal rate, as a function of the initial/resetting coordinate, shows both continuous and discontinuous transitions. Each of these cases are thoroughly analyzed.  Finally, we provide exact analytical results for the conditional escape times and the corresponding probabilities for the reset induced diffusive dynamics. The conditional escape times are found to display non-trivial behavior as the resetting rate is varied. It is noticed that resetting can induce a faster escape for the trajectories which start very close to an undesired boundary and escape through the further boundary. Looping transient trajectories, which usually cause large fluctuations in such cases, get curtailed by resetting hence a speed-up in the overall first passage time is observed.

Recent single-molecule experiments, including pulling studies on proteins and nucleic acid folding, and single-molecule fluorescence spectroscopy have raised a number of questions that stimulated theoretical and computational investigation on barrier-crossing dynamics \cite{BDB2015,BDB2017,Berezh06,Dagdug-03,Makarov-1,Makarov-2,Makarov-3,Makarov-4}. A model of chemical reactions considers
transitions of a Brownian particle between two deep wells of a
one-dimensional double-well potential, separated by a high
barrier. The fine structure of these trajectories has been analyzed in order to gain new insights into escape dynamics. Any trajectory of a diffusing particle making a transition between two end points of an interval in one dimension will have two components: the transition path segment -- part of the trajectory that leaves the starting point for the last time and goes to the end without returning to the starting point, and the looping segment where a number of loops that start and end at the same starting point are completed \cite{BDB2017}. We believe that the latter may be visualized as a non-instantaneous resetting process -- a class of process which assumes that resetting is a physical process, and thus reset/return to a preferred location will indeed take time \cite{HRS,non-inst-det-2,non-inst-det-3,non-inst-det-4,non-inst-det-5}. Bridging this gap can be useful to analyze, for example, the capture and escape of densely negatively charged C-terminal tails of cytosolic proteins in nanopores of $\beta$ -barrel channels, as well as artificial constructs assembled from albumin molecules and covalently attached negatively charged peptides mimicking C-termini of cytosolic proteins \cite{Rostovtseva} --
potential research avenues that need to be explored in future.




\section{Acknowledgements}
AP gratefully acknowledges the DST-SERB Start-up Research Grant Number SRG/2022/000080 and DAE, India for research funding. This study was partially supported by CONACyT under the grant Frontiers Science No. 51476.  

\appendix

\begin{widetext}
\section{Wide-to-narrow tube}\label{appendixa}
In the main text, we were working with narrow-to-wide conical tube whose radius varied as $R(x)=\lambda x+a$ with $\lambda\ge 0$. Here we give the results corresponding to wide-to-narrow tubes, {\it i.e.}, when the radius varies as $R(x)=\lambda (L-x)+a$ with $\lambda\ge 0$, in the case of one reflective and one absorbing boundary. In other words, the reflective boundary at $x=0$ is now wide, whereas the absorbing boundary at $x=L$ is narrow. The full expressions for this case
can be obtained by substituting $\lambda$ by $-\lambda$ and $a$ by $a+\lambda L$ in the expressions of Section \ref{sec:reflabs}.

We do not repeat the entire methodology here, but just summarize the main results of our interest. For instance, the first passage density of underlying process is given by
\begin{gather}
    \tilde{f}(s|x_0)=\frac{a}{(\lambda(L- x_{0})+a)}\frac{[-\lambda \sinh(ux_{0})+u(\lambda L+a) \cosh(ux_{0})]}{[-\lambda \sinh(uL)+u(\lambda L+a) \cosh(uL)]}.
\end{gather}
The first two moments of the first passage time distribution are given by
\begin{align}\label{eq:T1arwn}
    \langle T(x_{0})\rangle=\frac{L^2(2\lambda L+3a)}{6D_\lambda a}-\frac{x_0^2[\lambda(3L- x_0)+3a]}{6D_\lambda[\lambda (L-x_0)+a]},
\end{align}
and

\begin{align}\label{eq:T2arwn}
    \langle T^2(x_{0})\rangle=\frac{x_0^4[\lambda (5L-x_0)+5a]}{60D_\lambda^2[\lambda(L-x_0)+a]}-\frac{L^4(4\lambda L+5a)}{60D_\lambda^2a}+\frac{L^4(2\lambda L+3a)^2}{18D_\lambda^2a^2}-\frac{L^2x_0^2(2\lambda L+3a)[\lambda(3L- x_0)+3a]}{18D^2a[\lambda(L- x_0)+a]}.
\end{align}

For all values of the starting position and $\lambda$ as small as $0.2$, the MFPT  is significantly larger than in the narrow-to-wide configuration, as seen in Fig. \ref{fig:ref3} in comparison to Fig. \ref{fig:ref1}.  In the expanding case the trajectories are subject to an effective entropic force in the direction of the absorbing wall.

 \begin{figure*}[h!]
\centering

		\includegraphics[width=7.5cm]{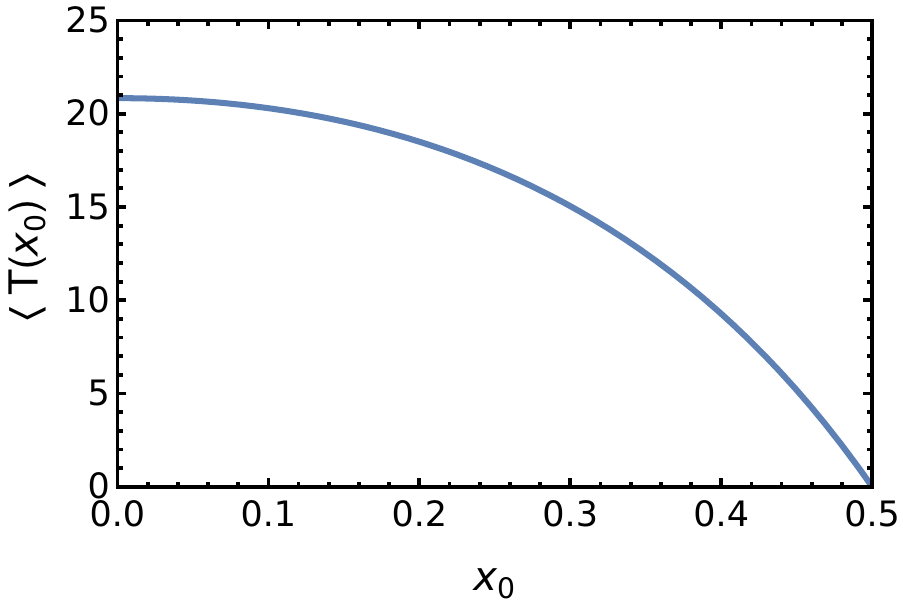}
		\label{fig:basic ref wn}
		\includegraphics[width=9cm]{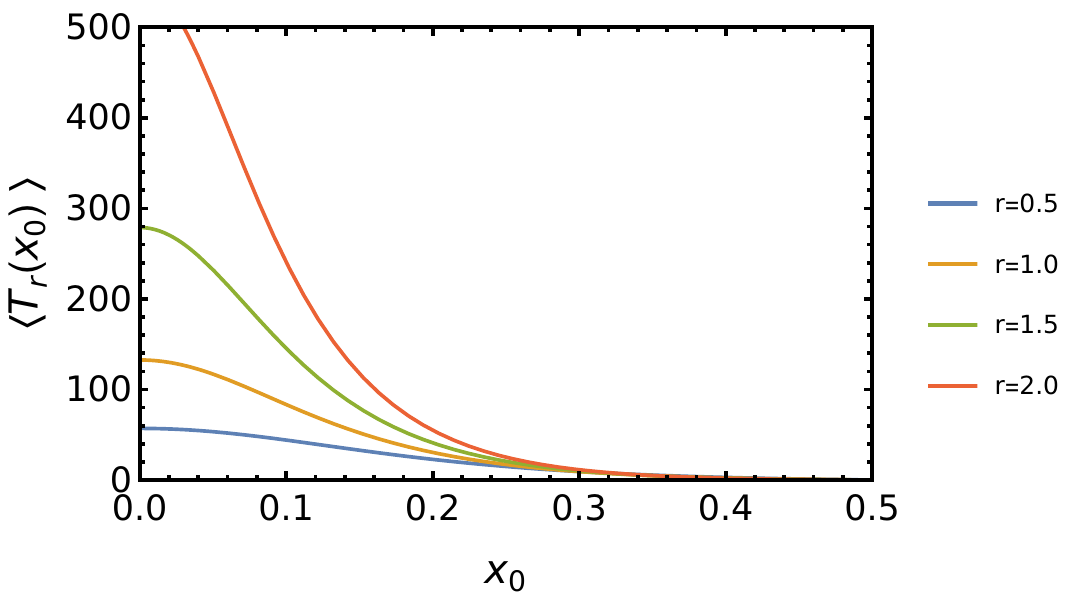}
		\label{fig:pos ref wn}

	\caption{Variation of MFPTs as a function of starting position $x_0$ without resetting (left) and with resetting (right) in the reflecting-absorbing case with a narrowing tube of parameters $a=0.1$, $D_\lambda=0.01$, $L=0.5$, $\lambda=0.2$.}
	\label{fig:ref3}
\end{figure*}

The first passage density under resetting is given in the Laplace domain by:
\begin{gather}
    \tilde{f}_{r}(s|x_0)= \frac{a(s+r)[-\lambda \sinh(\beta x_{0})+\beta (\lambda L+a) \cosh(\beta x_{0})]}{ra[-\lambda \sinh(\beta x_{0})+\beta (\lambda L+a) \cosh(\beta x_{0})]+s(\lambda(L- x_{0})+a)[-\lambda \sinh(\beta L)+\beta (\lambda L+a) \cosh(\beta L)]},
\end{gather}
from which we obtain the MFPT with resetting as
\begin{align}
    \langle T_{r}(x_{0})\rangle=\frac{(\lambda (L-x_{0})+a)[-\lambda \sinh(\alpha L)+\alpha (\lambda L+a) \cosh(\alpha L)]-a[-\lambda \sinh(\alpha x_{0})+\alpha (\lambda L+a) \cosh(\alpha x_{0})]}{ra[-\lambda \sinh(\alpha x_{0})+\alpha (\lambda L+a) \cosh(\alpha x_{0})]}.
\end{align}

\begin{figure*}[h!]
	
		\includegraphics[width=8.6cm]{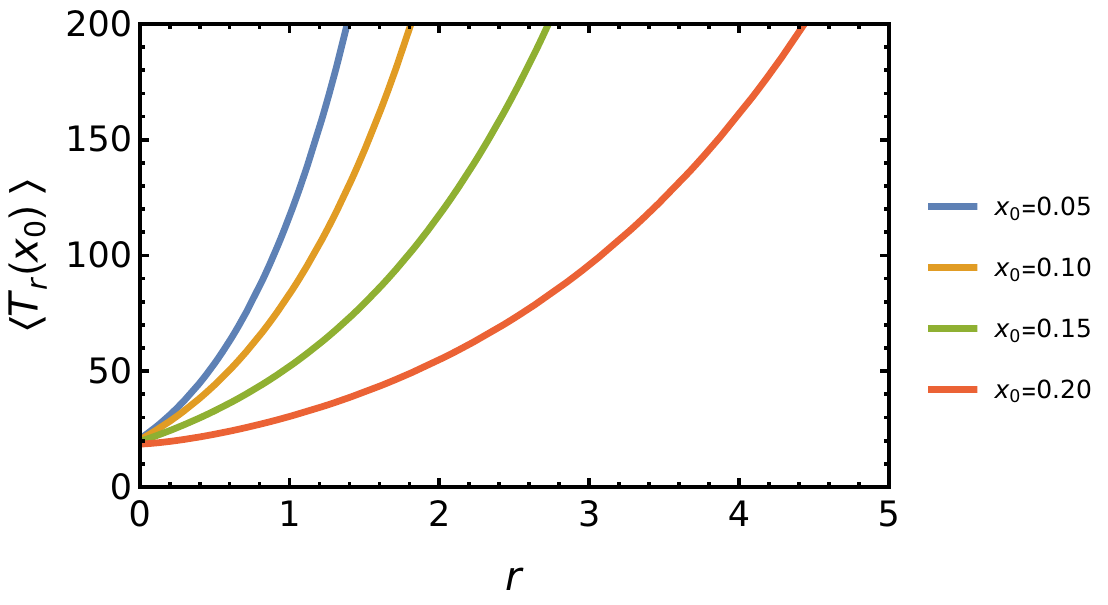}
		\includegraphics[width=8.5cm]{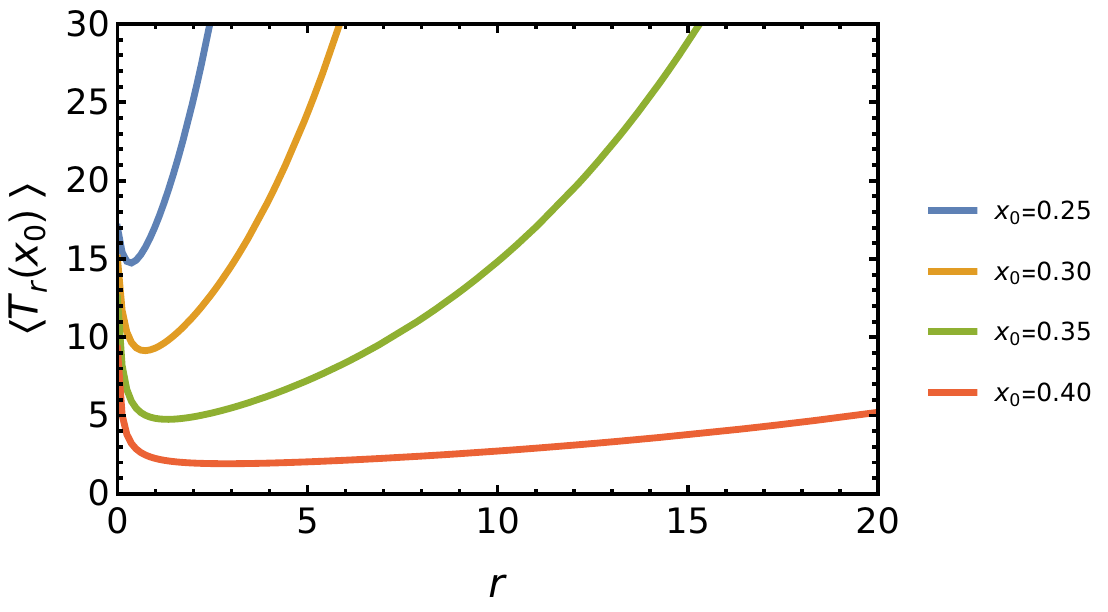}
	
	\caption{Variation of the MFPT as a function of the resetting rate for starting positions such that resetting is not useful (left) and useful (right), in the reflecting-absorbing case with a narrowing tube. The parameters are $a=0.1$, $D_\lambda=0.01$, $L=0.5$, $\lambda$=0.2. The criterion $CV=1$ gives a critical initial position $x_0^{c}$=0.2067.}
	\label{fig:ref4}
\end{figure*}
We see that for $v=x_0/L=0$, the criteria favouring resetting is never satisfied since $(\frac{2\lambda+3\overline{a}}{\overline{a}})^2 \leq \frac{-3(4\lambda+\overline{a})}{10\overline{a}}$.
In the example of Figure \ref{fig:ref4} (left), resetting increases the MFPT of all the initial conditions that are too close to the reflective wall ($x_0<0.2067...$), whereas an optimal $r^*>0$ exists for the initial conditions that are further away (right).
When $\lambda=0$, the criteria $CV>1$ reduces to
$    -25v^4+70v^2-31 \geq 0.
$

\section{Numerical simulations}
When running simulations, we considered an overdamped point-like  Brownian particle diffusing into the 3D conical tube, where the particle dynamics can be described by means of the Langevin equation, namely,
\begin{equation}
\frac{d\vec{r}}{dt}= \; \vec{\xi}(t),
\label{eq:lang}
\end{equation}
where $\vec{r}=(x,y,z)$. Consequently, the Brownian particle is subject to a Gaussian noise $\vec{\xi}(t)$ of zero mean and uncorrelated in time i.e., $\langle \vec{\xi}(t)  \rangle=\langle (\xi_{x}(t),\xi_{y}(t),\xi_{z}(t))\rangle=0$ and the auto-correlation functions are $\langle \xi_{i}(t),\xi_{j}(t') \rangle= 2D_0\delta_{ij} \delta(t-t')$ where $i,j=x,y,z$. When running simulations we took the time step $\Delta t=10^{-7}$ and the effective diffusivity $D_\lambda=0.01$, so that $\sqrt{2D_0 \Delta t} \ll 1$, and $k_B T$ is set equal to 1. Using $D_\lambda = D_0 / \sqrt{1+\lambda^2}$ and $\lambda=0.2$, we have a factor of $\sqrt{1.04}=1.0198$ between them. Stochastic averages were obtained as ensemble averages over $5.0 \times 10^{4}$ trajectories. Our simulation results have shown that
the reduction of axial diffusion for three-dimensional conical tubes with resetting, to the effective one-dimensional description in terms of the modified FJ equation, are in good agreement with a relative error of less than 3\%.

\end{widetext}

\end{document}